\documentclass[%
 reprint,
 amsmath,amssymb,
 aps,
]{revtex4-2}

\usepackage{hyperref}
\hypersetup{
    colorlinks=true, 
    linkcolor=cyan,
    citecolor=magenta, 
    filecolor=magenta, 
    urlcolor=cyan,
    runcolor=cyan
}
\usepackage{hyperref}
\usepackage{graphicx}
\usepackage{dcolumn}
\usepackage{bm}
\usepackage{subfig}
\usepackage{color,soul}
\usepackage{float}
\usepackage{mathtools}
\usepackage{siunitx}
\usepackage{booktabs}

\usepackage[english]{babel}

\makeatletter
\renewcommand{\fnum@figure}{FIG. \thefigure}
\makeatother

\begin{document}

\title{Machine learning-based analysis of experimental electron beams and gamma energy distributions}


\author{M. Yadav$^{1,2,3}$}
\email{yadavmonika@g.ucla.edu}

\author{M. Oruganti$^{1}$}
\author{S. Zhang$^{1}$}
\author{B. Naranjo$^{1}$}
\author{G. Andonian$^{1}$}
\author{Y. Zhuang$^{4}$}
\author{{\"O.} Apsimon$^{2,3}$}
\author{C. P. Welsch$^{2,3}$}
\author{J. B. Rosenzweig$^{1}$}

\affiliation{
$^1$Department of Physics and Astronomy, University of California Los Angeles, California 90095, USA
}
 \affiliation{ 
 $^2$ Cockcroft Institute, Warrington WA4 4AD, UK}
\affiliation{ 
 $^3$Department of Physics, University of Liverpool, Liverpool L69 3BX, UK}
 \affiliation{ 
 $^4$ Carnegie Mellon University, Pennsylvania 15213, USA}

\date{\today}
\begin{abstract} 

The photon flux resulting from high-energy electron beam interactions with high field systems, such as in the upcoming FACET-II experiments at SLAC National Accelerator Laboratory, may give deep insight into the electron beam's underlying dynamics at the interaction point. Extraction of this information is an intricate process, however. To demonstrate how to approach this challenge with modern methods, this paper utilizes data from simulated plasma wakefield acceleration-derived betatron radiation experiments and high-field laser-electron-based radiation production to determine reliable methods of reconstructing key beam and interaction properties. For these measurements, recovering the emitted 200 keV to 10 GeV photon energy spectra from two advanced spectrometers now being commissioned requires testing multiple methods to finalize a pipeline from their responses to incident electron beam information. In each case, we compare the performance of: neural networks, which detect patterns between data sets through repeated training; maximum likelihood estimation (MLE), a statistical technique used to determine unknown parameters from the distribution of observed data;
and a hybrid approach combining the two.  Further, in the case of photons with energies above 30 MeV, we also examine the efficacy of QR decomposition, a matrix decomposition method. The betatron radiation and the high-energy photon cases demonstrate the effectiveness of a hybrid ML-MLE approach, while the high-field electrodynamics interaction and the low-energy photon cases showcased the machine learning (ML) model's efficiency in the presence of noise. As such, while there is utility in all the methods, the ML-MLE hybrid approach proves to be the most generalizable.

\end{abstract}

\maketitle

\section{Introduction}

The transverse acceleration of very energetic charged particles provokes electromagnetic radiative processes, generally described as variants of synchrotron radiation. However, in specific cases encountered in modern applications, such as those found in advanced accelerator research, the radiation is more specifically referred to by the type of external force giving rise to the relevant acceleration. For instance, for electron beams in plasma wakefield accelerators (PWFA), the transverse restoring force due to the presence of an uncompensated ion distribution \cite{jamie_blowout} gives rise to periodic betatron motion. Hence, the radiation produced in this process is called betatron radiation \cite{Sebastian_2013}. In the case of a high-energy electron beam's interaction with a sufficiently intense laser beam, we reach the non-perturbative regime of strong-field QED (SFQED) \cite{di2018implementing}, observing effects such as nonlinear inverse Compton scattering (NICS).  

These radiative processes are of high importance in advanced accelerator experimentation. Gamma rays emitted from the betatron radiation process in under-dense plasma wakefields or by NICS produce unique experimental signatures that reveal interaction physics at the challenging ultra-short spatial and temporal experimental scales. Additionally, beam-based diagnostics minimize interference with extreme high-field environments relevant to advanced acceleration techniques. Recent discussions illustrated by published work indicate that the user community in the first phase of FACET-II research will be dominantly concerned with measurements centered on relatively large emittance and higher energy beam processes, which produce spectra typically localized in the 0.5 to 10's of MeV spectral range \cite{san2019betatron}.

Betatron radiation and related processes are thus of interest as they provide uniquely powerful probes of high-energy beam-plasma and beam-radiation interactions. They arise, as noted above, in two most relevant ways: in the case of betatron radiation, the oscillations arise from strong focusing fields with a nearly linear offset from the axis. In the case of intense laser-electron interactions, the radiation is closely related to undulator radiation, with the oscillation driven by nearly constant electromagnetic wave fields. In both cases, approximately sinusoidal transverse motion gives rise to radiation with well-understood properties. 

For betatron radiation, the focusing arises from nominally linear fields due to an approximately uniform ion channel formed in a blowout regime \cite{jamie_blowout} of PWFA, as discussed further below. An important caveat exists here - the electron beams in next-generation experiments at FACET-II may be dense enough to provoke ion motion\cite{ioncollapse}, and a more detailed treatment is demanded in this case\cite{weiming}. Concentrating for the moment on the simpler case of linear oscillations in the uniform ion-column limit, the oscillation wavelength can be written as follows:

\begin{equation}
    \lambda_\beta = \lambda_p\sqrt{2\gamma},   
\end{equation}
where $\lambda_p=2\pi/k_p $ is the density-dependent plasma wavelength, with
\begin{equation}
    k_p=\sqrt{\frac{e^2n_0}{\epsilon_0m_ec^2}}.
\end{equation}

In linear betatron oscillations, the amplitude $x_\beta$ is related to the so-termed \textit{undulator parameter} $K=\gamma k_\beta x_\beta$, which is in turn related to the on-axis, resonant radiation wavelength in the undulator limit ($K\leq 1$) as follows:

\begin{equation}
    \lambda_r = {\frac{\lambda_\beta}{2\gamma^2}}\left(1+\frac{K^2}{2}\right)    
\end{equation}

For $K$ notably greater than unity, the radiation pattern does not show a peak near this resonance and its harmonics, but is more effectively described as a synchrotron-like spectrum.  At FACET-II, the plasma wavelength $\lambda_p$ ranges from 3-300 $\mu$m (plasma density $n_0$ ranging from $10^{16}$ to $10^{20}$ cm$^{-3}$), while $\gamma$ ranges from 2000 to 25000.  

In wakefield accelerators, there are two beam bunches: the drive, which provides the primary excitation of the wake, and the witness, a typically lower charge accelerating bunch. For the drive beam, particularly during FACET-II commissioning, the spectrum is notably different, as the beam may have many trajectories with $K\gg 1$, due to the sizeable transverse emittance and concomitant beam size. As a result, the emitted spectrum is, as noted above, similar to that of synchrotron radiation \cite{Sebastian_2013} from bend magnets, with a characteristic maximum near the critical energy $\hbar\omega_c$ associated with the minimum bend radius of the motion. This synchrotron-like spectrum is broad and, therefore, has a significantly low energy component that cannot be ignored in the radiation environment.

As discussed above, in blowout regime PWFA, a dense drive beam generates a linear focusing force by repelling the plasma electrons away from its path, while the plasma ions, being much more massive, are nominally left uniformly distributed inside the blowout bubble. The majority of drive beam electrons will oscillate in this bubble, In a witness beam subject to this focusing force in plasma, nearly electrons undergo harmonic transverse betatron oscillations at the spatial frequency $k_\beta\simeq \sqrt{2\gamma}k_p$, giving rise to betatron radiation. Because information about the properties of the beam (\textit{e.g.}, emittance, energy) is encoded in the betatron radiation spectrum, measurements of this radiation can be used in principle to reconstruct beam parameters, mainly through the changing characteristics of the radiation with strength parameter $K$. A preliminary study has examined the possibility of using betatron radiation to reconstruct the accelerated electron beam transverse profile \cite{tracespace}. 

One method of extracting information about beam parameters from such radiation measurements is maximum likelihood estimation (MLE). This statistical technique determines unknown parameters from a given distribution of observed data. Various beam diagnostic devices and techniques already exist, such as the beam current transformer used to measure beam  charge and laser-Compton scattering used to measure transverse beam spot size and emittance \cite{koziol2001beam}. In addition, machine learning (ML) methods are also utilized for different implementations of beam diagnostics and controls \cite{fol2019application}. With new instruments being developed for a rich range of experimental scenarios in FACET-II and other advanced accelerator labs worldwide, it is necessary to quantify the effectiveness of these techniques for extracting relevant information from radiative signals.  This work aims to assess the ability of MLE and machine learning to extract beam and radiation parameters from measurements of betatron radiation and related processes (i.e., laser scattering).

There are many such advanced acceleration experimental efforts, in particular aimed at PWFA development, which has already demonstrated acceleration gradients over 50 GeV/m \cite{nature50gevm}, with other scenarios identified that can extend this gradient to the TeV/m range \cite{tevpermeter,Manwani2020}. There is considerable research interest in developing the PWFA aimed at high gradient acceleration applications such as linear collder and X-ray free-electron laser (XFEL). As such, significant new facilities are now being commissioned to explore the physics issues supporting PWFA research as well its near-term application, such as the FACET-II mentioned above \cite{facet2}, and FLASHForward \cite{flashforward} at DESY (Hamburg). The EuPRAXIA initiative at the SPARC Lab\cite{Pompili2022} in Frascati, Italy, will be the first to proceed to dedicated application as an XFEL user facility.

As high-energy electrons interact, their emitted gamma radiation has been effectively used to determine aspects of the state of the original particles and the characteristics of the interactions. For FACET-II, one must signicantly extend to the effectiveness of these approaches in new experimental environments. One will at FACET-II, confront photon energies across a wide range, as described above. For the spectral region ranging from 200 keV through 30 MeV, a dedicated Compton spectrometer, termed CPT, is used to measure the distribution of photons in energy directly through their "conversion" to Compton electrons. In addition, one angular dimension (vertical) along a lithium Compton converter target, placed at the entrance of a horizontally dispersive magnet spectrometer is decoupled from the energy measurement; a doubledifferential angular-energy spectrum (DDS) is thus obtainable. The information gleaned from DDS may give more sophisticated information concerning the beam and interaction deteils.   The spectrometer's magnetic field geometry is sextupole-like, permitting the needed sizeable single-shot energy range. To determine the energy of the photons emitted at energy above 30 keV, where to the photon-driven spectrum from the converter target changes from Compton scattering to pair-production dominated, a new pair spectrometer (Positron-Electron Detector for Radiative Observations, PEDRO) is used. It also similarly utilizes a converter target composed of beryllium, in tandem with a magnet array more appropriate for higher energy electron (and positron) measurements. Because photons of interest particularly in the SFQED experiment will have energies between 10 MeV and 10 GeV, they will produce copious electron-positron pairs through the nuclear field interaction. Under the influence of a magnetic field, highly forward-directed electrons and positrons bend in opposite directions and at different energy-dependent angles. The  An array of detection cells measure the number of electrons or positrons that strike at the full range of relevant locations. This information can be employed straightforwardly in the case of single-gamma detection to establish the incoming photon energy. In the case of a very large number simultaneous gamma rays incidents on the detector at a distributin of energies, the analysis is significantly more challenging. 

To analyze these experimental issues in detail, we proceed as follows. In Sec. \ref{sec:MLE_2}, we introduce the analytical description and main concepts behind MLE and lay out a method for using an MLE algorithm to identify beam parameters from different types of betatron radiation data. In Sec. \ref{sec:ML}, we lay out a method for using neural networks to accomplish the same task of identifying beam parameters from radiation data. In Sec. \ref{sec:reconstruction}, we discuss spectral reconstruction for a FACET-II Compton spectrometer using an EM algorithm and ML methods. In Sec. \ref{sec:Pairspec}, we discuss methods specific to the pair spectrometer analysis, including ML, MLE, and QR decomposition (wherein the detector response matrix is decomposed into two smaller, special matrices) methods. Finally, we offer conclusions and discuss the outlook for future work in Sec. \ref{sec:conclusion}.

\section{Reconstruction of Beam Parameters Using MLE}
\label{sec:MLE_2}

To introduce the analysis methods, we first discuss some preliminary considerations on the approaches taken. Beginning with MLE, generally stated, the goal of the analysis is to determine the value of an initially unknown parameter $\sigma$ which allows the probability distribution function $f(x|\sigma)$ to best model a set of observed $x$ vectors. This is accomplished by calculating a likelihood function $L(\sigma|x)$, which is related to the probability distribution function by $L(\sigma|x)=f(x|\sigma),$ where $L(\sigma|x)$ specifies the likelihood of $\sigma$ given $x$. Given a set of $N$ observations of $x$ vectors, the overall likelihood is given as the product of the likelihoods of each $x$ vector \cite{MYUNG200390} as given in Eq. \ref{eq:likelihood}.

\begin{equation}
    L(\sigma|x_1, x_2, ... , x_N)= \prod_{n=1}^{N} L(\sigma|x_n)
     \label{eq:likelihood}
\end{equation}

The value for the parameter $\sigma$, which is most likely to have produced a given set of observed data, is then determined by maximizing the likelihood $L(\sigma|x_1, x_2,..., x_N)$ concerning $\sigma$. In practice, maximizing the natural log of the likelihood, known as the log-likelihood, is often referred to as maximizing the linear likelihood because, while these two tasks are essentially equivalent, working with log-likelihood avoids possible problems with arithmetic underflow \cite{bishop}. 
\begin{equation}
    \ln L(\sigma|x_1, x_2, ... , x_N)= \sum_{n=1}^{N} \ln L(\sigma|x_n)
     \label{eq:sumrad}
\end{equation}
Therefore, in this work, we use the log-likelihood computed using Eq. \ref{eq:sumrad}.

To provide a set of physical predictions concerning experimental outcomes in blowout regime PWFA measurements, we have developed a betatron radiation simulation code that tracks particles through externallly-given fields and computes radiation using Liénard–Wiechert potentials. This custom tracking code is written in C++ and parallelized using MPI. In the algorithm, macro-particles are first randomly sampled from a simulated beam distribution for the experimental conditions in question. The extracted particles are then tracked, in the simplest model, through idealized acceleration and focusing fields using a 4th-order Runge-Kutta (RK4) method. The fields consist of linear focusing forces and a constant accelerating force. We note that the externally-derived trajectories may also be obtained from particle-in-cell codes, to obtain a detailed picture of a given experimental case. To illustrate our methodology, however, we concentrate on the idealized case. Finally, the particle trajectories obtained in these varietly of methods are used to evaluate the Liénard–Wiechert integral numerically. It provides a model for examining many relevant scenarios encountered in the FACET-II experimental PWFA program. This method is applied principally to CPT Compton-based spectrometry, in which it is much more straightforward to obtain the primary photon spectrum. In the case of higher energy interactions (mainly for the strong-field QED experiments), the main challenge is to obtain the spectrum itself, not, \textit{e.g.}, derived beam parameters per se. 

\begin{figure}[h!]
    \centering
    \includegraphics[width=\columnwidth]{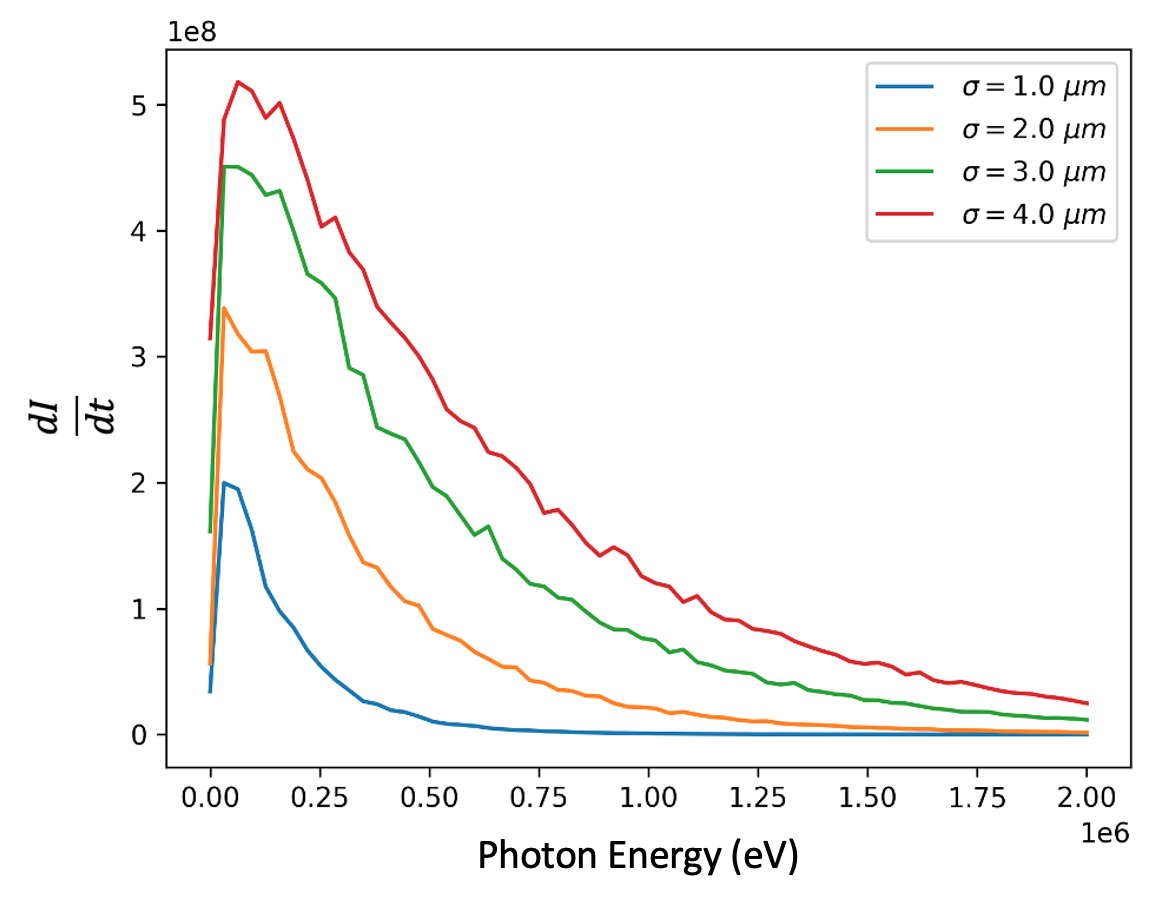}
    \caption{Radiation spectra, generated by the custom tracking code, for several spot sizes.}
    \label{fig:raw_spec_ex}
\end{figure}

\begin{table}
\caption{\label{tab:sim_parameters}Simulation parameters used for generating a set of radiation spectra to test the MLE algorithm. }
\begin{ruledtabular}
\begin{tabular}{cc}
Parameter & Value \\
\hline
$\varepsilon_x, \varepsilon_y$ $\si{(\mu m)}$ & 3  \\
ion charge state & 1 \\
$L$ $\si{(cm)}$& 5  \\
$E_{\mathrm{accel}}$ $\si{(V/m)}$ & $8\times10^9$  \\
$E$ $\si{(GeV)}$ & 10  \\
$\epsilon_{\mathrm{min}}$ $\si{(MeV)}$ & 0  \\
$\epsilon_{\mathrm{max}}$  $\si{(MeV)}$ & 2 \\
$\epsilon$ points & 64\\
$\epsilon$ grid spacing & linear  \\
$\phi_x$ min $\si{(mrad)}$  & -.001 \\
$\phi_x$ max   $\si{(mrad)}$ & .001 \\
$\phi_x$ points & 64 \\
$\phi_x$ grid spacing & linear \\
$\phi_y$ $\si{(rad)}$  & 0 \\
steps per $\si{m}$  $\si{(m^{-1})}$ & 40,000 \\
\hline  
\end{tabular}
\end{ruledtabular}
\end{table}

The radiation contribution from each particle is given by the Liénard–Wiechert integral

\begin{equation} \label{eq:lw1}
 \bm{V}_i =  \int_{t_i}^{t_f} \frac{\bm{n} \times (( \bm{n} - \bm{\beta} ) \times \dot{\bm{\beta}} )}{(1 - \bm{n} \cdot \bm{\beta})^2} e^{i \omega ( t - \bm{n} \cdot \bm{r}(t) / c )} dt
\end{equation}

\noindent where $\bm{n}$ is the average vector in the direction of the radiation, $\bm{r}$ and $\bm{\beta}$ are the normalized position and velocity, respectively, which are computed by the particle tracking code, and $\omega$ is the angular frequency of the radiation. Finally, the radiation contribution from each particle is summed through the superposition of the fields, as given in Eq.~\ref{eq:lw2}, to give the double differential radiation spectrum.

\begin{equation} \label{eq:lw2}
\frac{d^2 I}{d\Omega d\epsilon} = \frac{e^2}{16 \pi^3 \epsilon_0 \hbar c} \left | \sum_i \bm{V}_i \right |^2
\end{equation}

\noindent In order to compute the spectrum, this expression is evaluated over a grid of different values of $\phi_x$, $\phi_y$, and $\epsilon$, and the spectrum is computed by integrating over $d\Omega = d\phi_x d\phi_y$. Because evaluations of the radiation intensity spectrum contributions are independent, and $\bm{V}_i$ add linearly, there are ample opportunities to parallelize the code.

The first task for this work is to correctly identify a beam's spot size from its radiation spectrum using MLE. First, the custom tracking code obtains several radiation spectra produced by beams with various spot sizes. Table \ref{tab:sim_parameters} displays the simulation parameters used. Next, the MLE compares these to some test spectra. Examples of such spectra, which demonstrate the difference in radiation scenarios generated by the beams with different radii, are shown in Fig. \ref{fig:raw_spec_ex}. 

A known test spot size is then arbitrarily chosen, and an additional radiation spectrum is obtained for the chosen beam. In this case, the test spot size is chosen to be 2 $\mu m$. Finally, the MLE algorithm uses previously generated training data to calculate a probability distribution that aims to predict the test spot size. To accomplish this, previously generated training sets of spectra are converted into probability distributions as a function of photon energy. Altogether this process forms a probability distribution function $f(x|\sigma)$, where $x$ is photon energy and $\sigma$ is spot size. This was done using Eq. \ref{eq:probability}, which treats the photon energy as a discrete variable, such that, for a spectrum with a number, $J$, of discrete photon energies. 

\begin{equation}
    f(x|\sigma)= \frac{dI}{dt}(x|\sigma) \div \sum_{j=1}^{J}\left(x_{j} \frac{dI}{dt}|\sigma \right)
    \label{eq:probability}
\end{equation}
 
\begin{figure}[h!]
    \centering
     \includegraphics[width=\columnwidth]{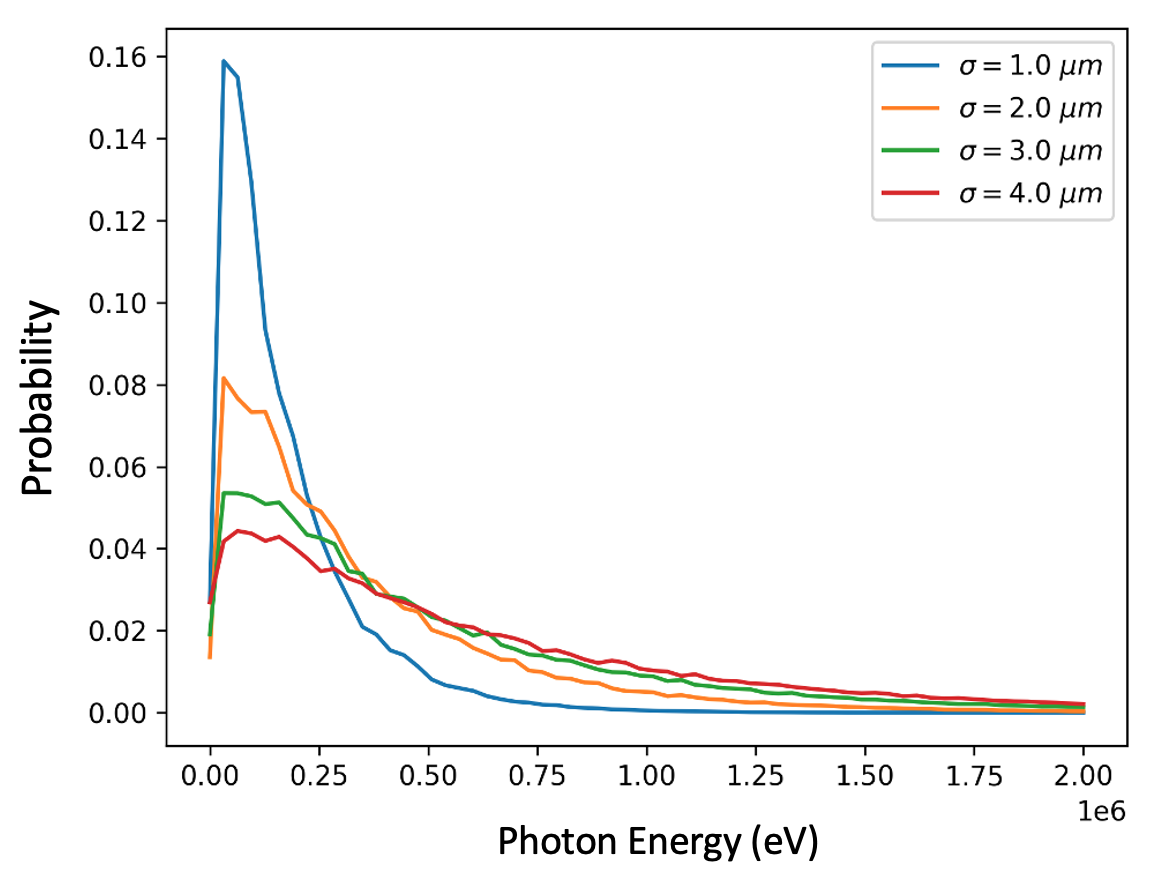}
    \caption{Radiation spectra, for several spot sizes, generated by the custom tracking code and converted to probability distributions, $f(x|\sigma)$, using Eq. \ref{eq:probability}.}
    \label{fig:prob_spec_ex}
\end{figure}

The resulting probability distributions are shown in Fig. \ref{fig:prob_spec_ex}.

\begin{figure}[h!]
    \centering
    \includegraphics[width=\columnwidth]{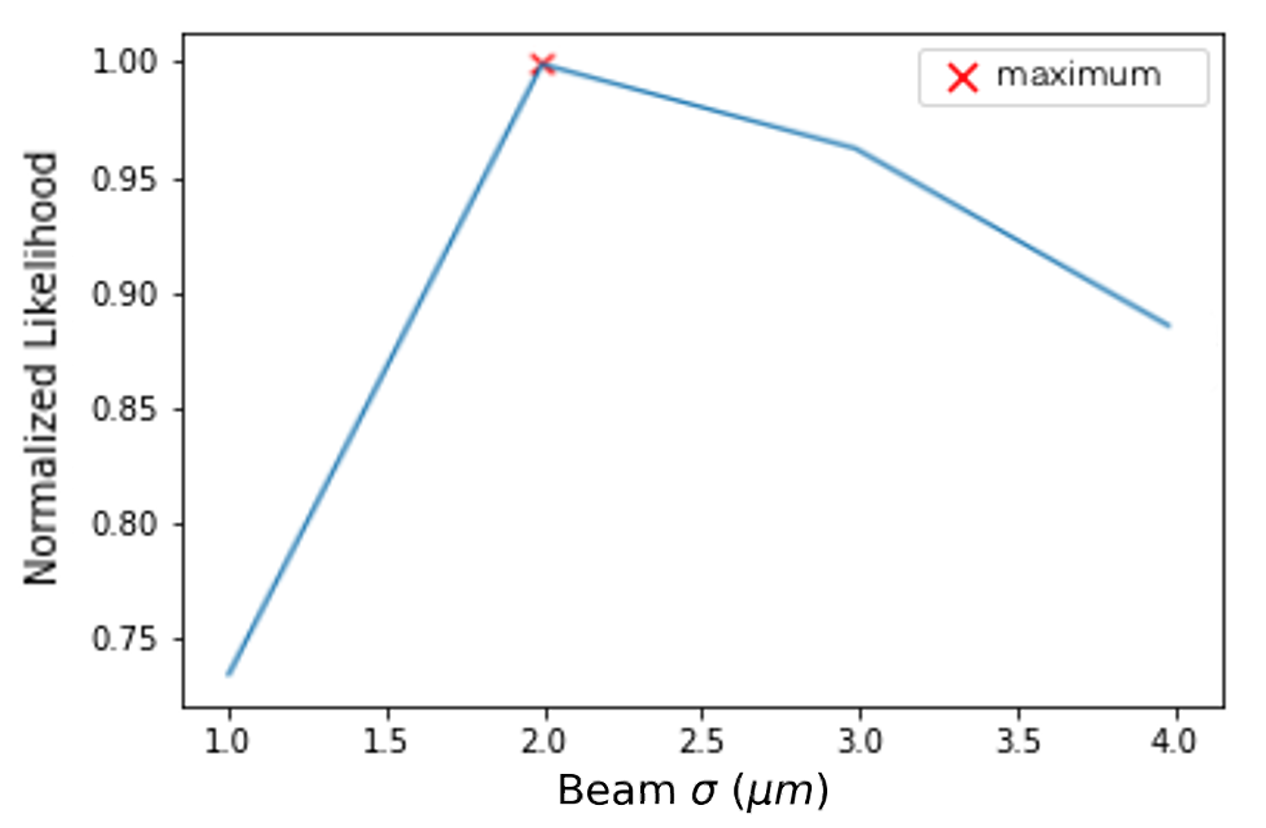}
    \caption{The normalized likelihood function reaches a maximum at the test spot size, 2 $\mu m$, which means that it has correctly identified the test spot size.}
    \label{fig:mle_example}
\end{figure}

When applying Eq. \ref{eq:likelihood} to photon energy spectra, each $x$ vector represents an observed photon with some energy value. These inputs are extracted from the test data. In authentic MLE, the test spectrum's $y$-axis should be in terms of a concrete number of objects (\textit{i.e.,} photons) but, because the scale of the test function only changes the scale of the calculated likelihood function and does not affect MLE results, this work also converts the test spectrum to a probability distribution $f_\mathrm{test}(x)$ using Eq. \ref{eq:probability} for ease of comparison. The likelihood that the probability distribution $f(x|\sigma)$ models the test spectrum $f_\mathrm{exp}(x)$ for different values of $\sigma$, can then be calculated by Eq. \ref{eq:sumrad}. That is, for a test spectrum of discrete photon energies $x_1, x_2, ..., x_J$.

\begin{equation}
    \label{final_llh}
    \ln L(\sigma|x)= \sum_{j=1}^{J} f_\mathrm{exp}(x_j)\ln f(x_j|\sigma)
\end{equation}

With the probability distributions in Fig. \ref{fig:prob_spec_ex} and the test spectra as inputs into the likelihood function, the likelihood can be plotted for spot size. For clarity, the likelihood functions in this work are plotted with the likelihoods normalized concerning the maximum value. In Fig. \ref{fig:mle_example}, the model is shown to correctly identify the test spot size of 2 $\si{\mu m}$.

\begin{figure}[h!]
    \centering
    \subfloat[Double differential distribution]{\includegraphics[width=\linewidth]{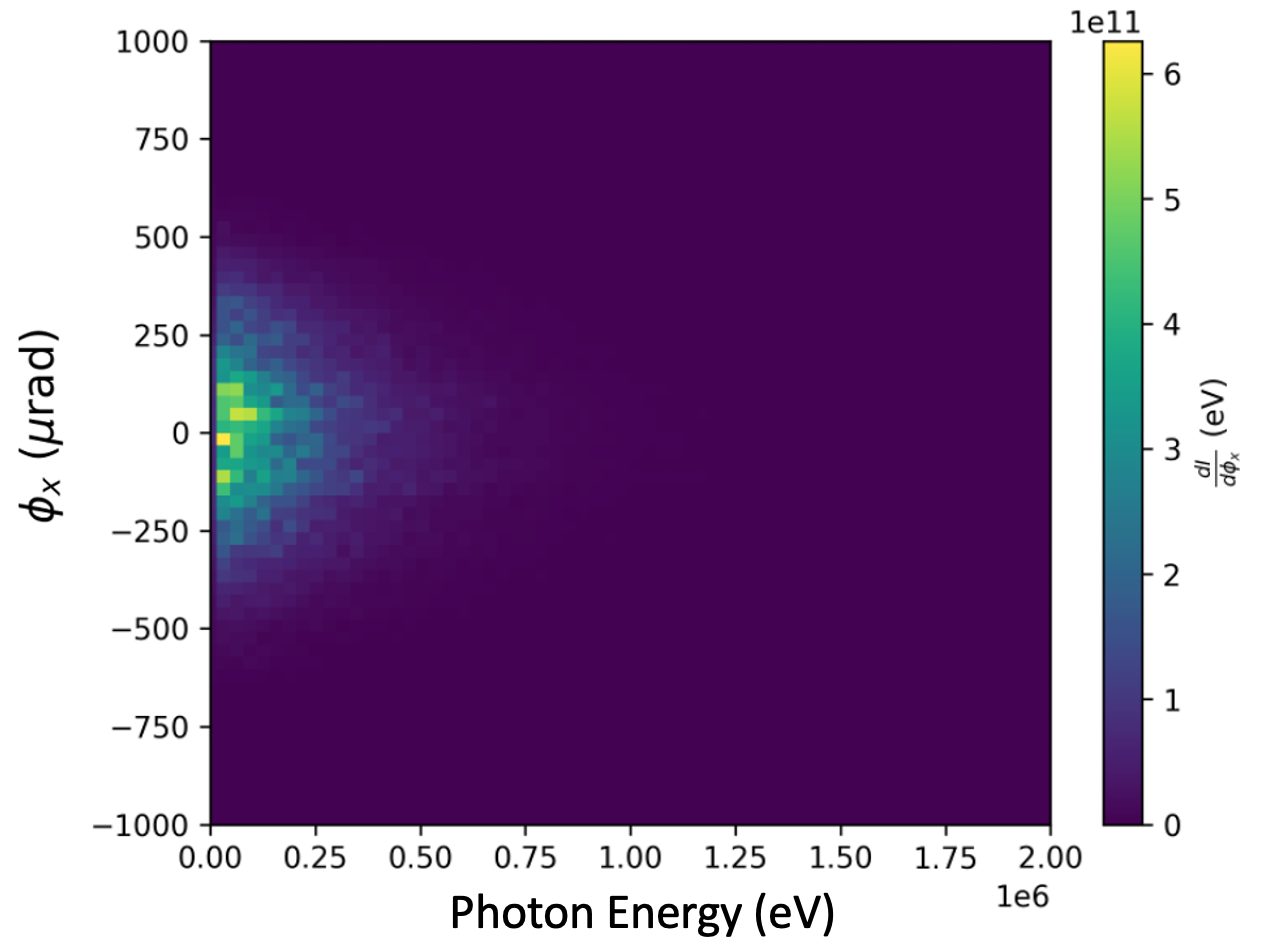}}
    \\
    \subfloat[Angular distribution]{\includegraphics[width=\linewidth]{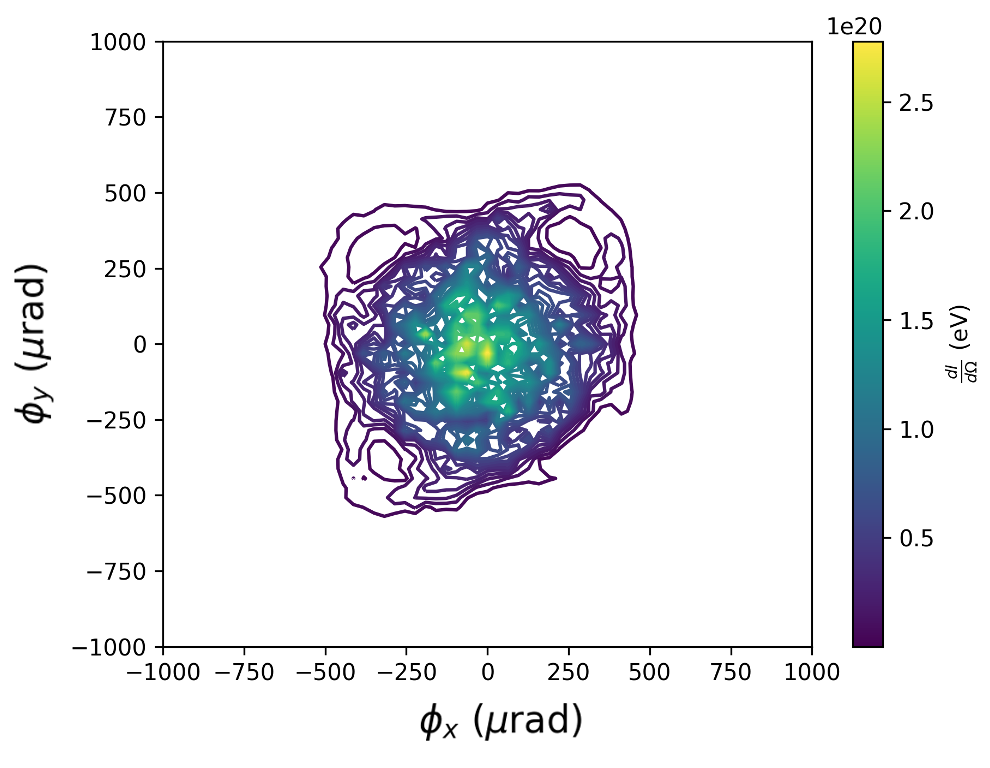}}
    \caption{Two types of 2D distributions plotted from the same radiation data. The radiation data was obtained through simulation using the parameters in Table \ref{tab:sim_parameters} and an arbitrary spot size of 1 $\mu m$.}
    \label{fig:2d_plots}
\end{figure}

After determining that this MLE model can successfully identify beam spot size using the 1D radiation spectrum, various simulation parameters are varied to determine their effect on the accuracy of MLE predictions. Overall, simulations with more resolved energy and angular distribution spectra produced plots that are easier to differentiate based on spot size, even by visual inspection, resulting in more robust MLE results. In particular, simulations with more resolved angular distributions also produced less noisy energy spectra, improving the accuracy of the MLE for energy spectra. In addition, simulating with a more significant number of particles also improved the MLE accuracy by reducing the effect of randomness in the simulations on the MLE results. However, both the resolution and number of particles were limited by simulation time and computing power, so the accuracy and precision of the MLE are at present ultimately limited by practical constraints. Nevertheless, the MLE can correctly identify spot size with precision to a tenth of a micron with these adjustments.  

\begin{figure}
    \centering
  \includegraphics[width=\columnwidth]{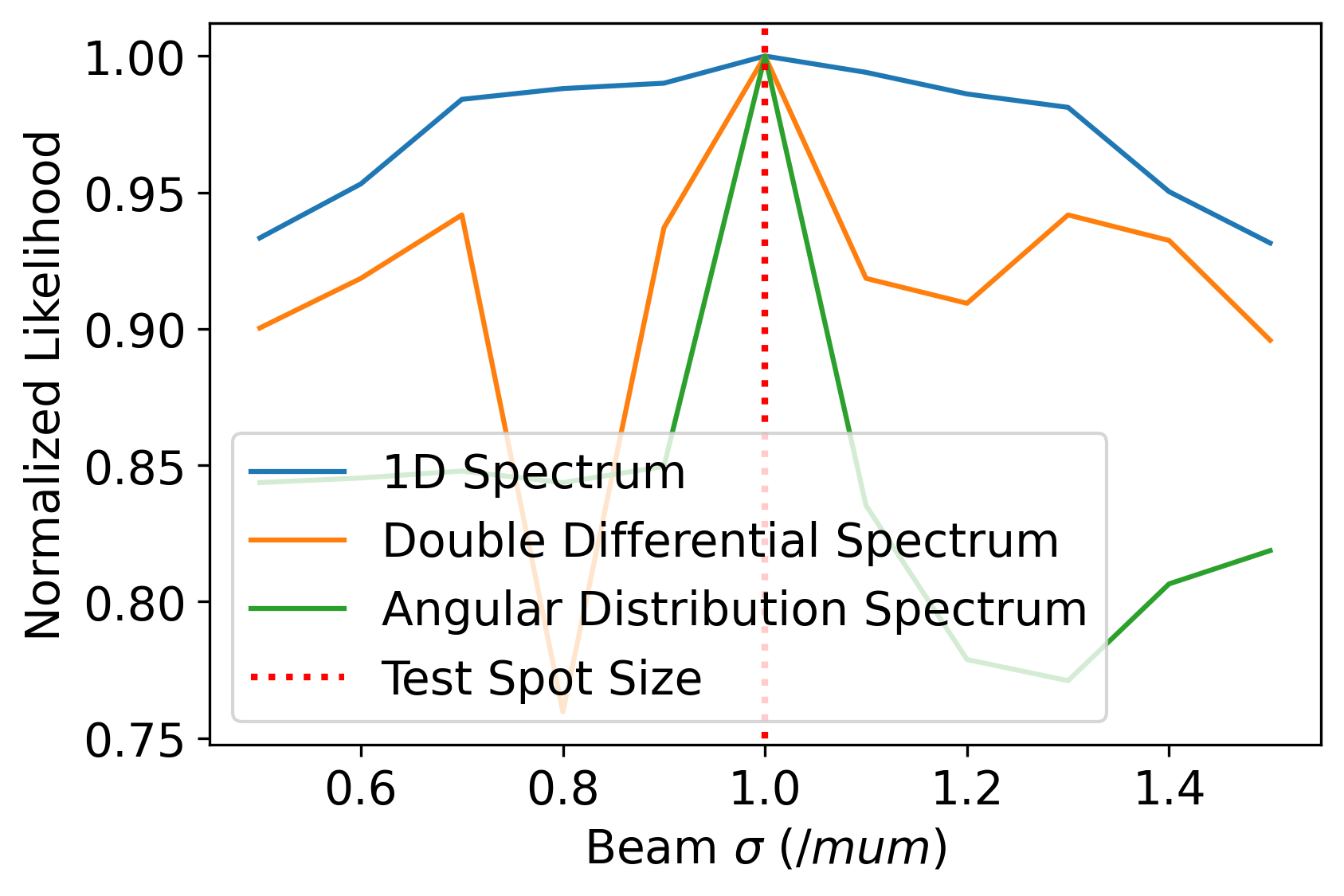}
    \caption{Likelihood functions for identifying a test spot size of 1 $\mu m$ from three types of radiation data: the 1D energy spectrum, double differential distribution, and angular distribution. Regardless of the type of radiation data used, the MLE algorithm successfully identifies the test spot size.}
    \label{fig:2d_mle}
\end{figure}
 
This same MLE model can be extended to analyze data for two-dimensional distributions, such as DDS distributions and 2D angular ($\theta_x,\theta_y$) spectral distributions, which . Examples of these plots are shown in Fig. \ref{fig:2d_plots}. The previously analyzed spectral type, as shown in Fig. \ref{fig:raw_spec_ex}, will henceforth be referred to as 1D radiation spectra to differentiate the three spectrum types.

In the case of two-dimensional distributions, MLE works in much the same way. The expression in Eq. \ref{final_llh} is summed over all points in the distribution, just as in the one-dimensional case. The results of MLE using two-dimensional distributions and the results of MLE using 1D spectra plotted from the same radiation data are shown in Fig. \ref{fig:2d_mle}. While both methods can correctly identify the test spot size of 1 $\si{\mu m}$, both likelihood functions have irregular shapes when compared to the likelihood function that uses the 1D energy spectrum, which is produced with the same radiation data. This artifact may cause problems in the future when the MLE becomes more complex or is desired to be more precise.

\begin{figure}
    \centering
    \includegraphics[width=\columnwidth]{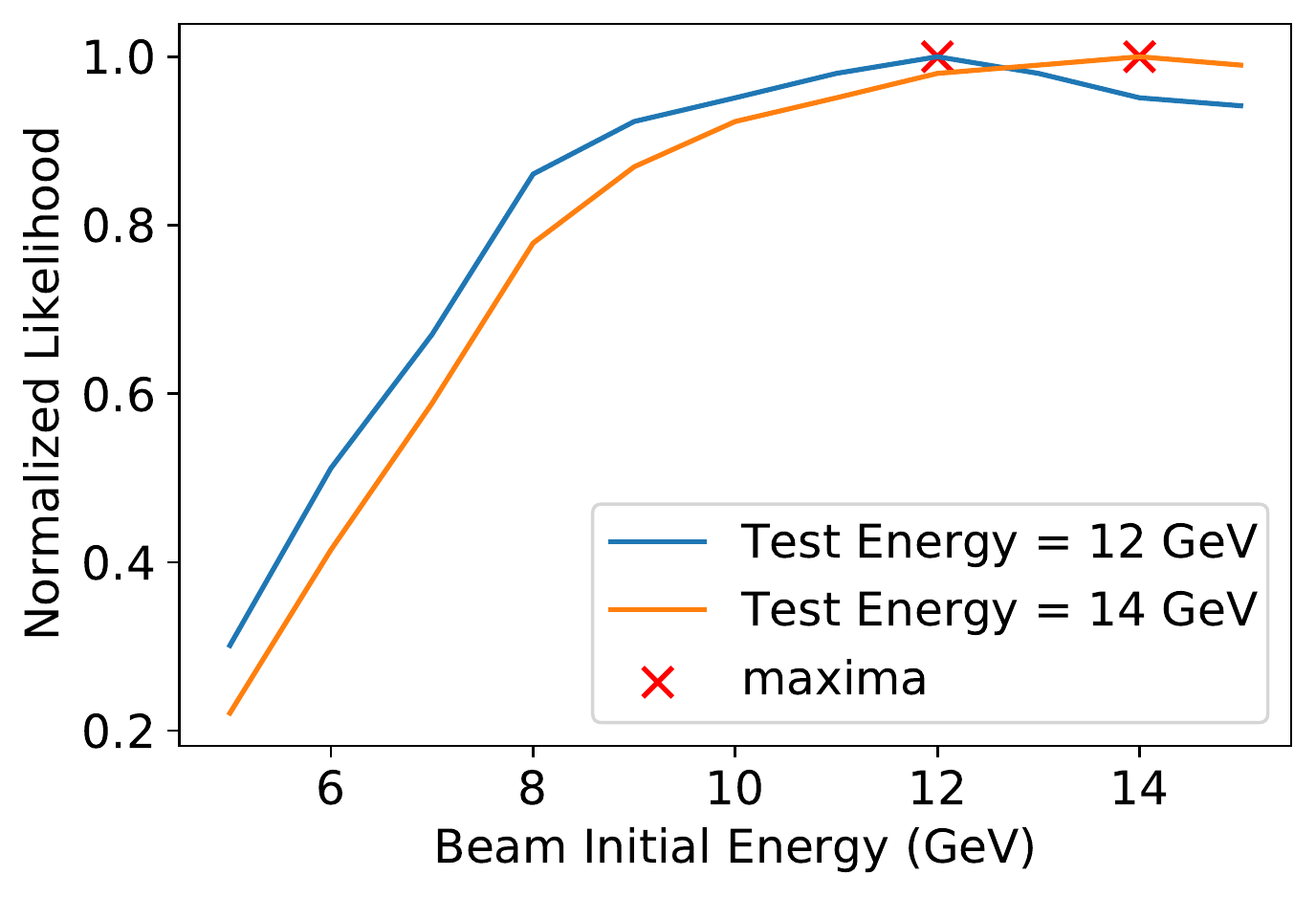}
    \caption{Likelihood functions for identifying two different initial beam energies, 12 and 14 GeV, using 1D energy spectra. The MLE algorithm successfully identifies both beam energy values.}
    \label{fig:energy_mle}
\end{figure}

In addition to being expanded to different distribution types, the MLE algorithm can also be expanded to identify different beam parameters, such as beam energy and emittance. In a basic test, the MLE algorithm was used to identify beam energy by running several simulations with only beam initial energy varied between them and replacing the spot size $\sigma$ with beam energy $\epsilon$ in Eq. \ref{final_llh} to obtain a likelihood function. In order to demonstrate the effectiveness of the MLE, two different test simulations were run: one with an initial beam energy of 12 GeV and another with 14 GeV. The log-likelihood functions for these two test beam energies are displayed in Fig. \ref{fig:energy_mle}. These two cases together show that MLE can be used to accurately identify a beam's initial energy given its betatron radiation energy spectrum. 

\begin{figure}
    \centering
    \includegraphics[width=\columnwidth]{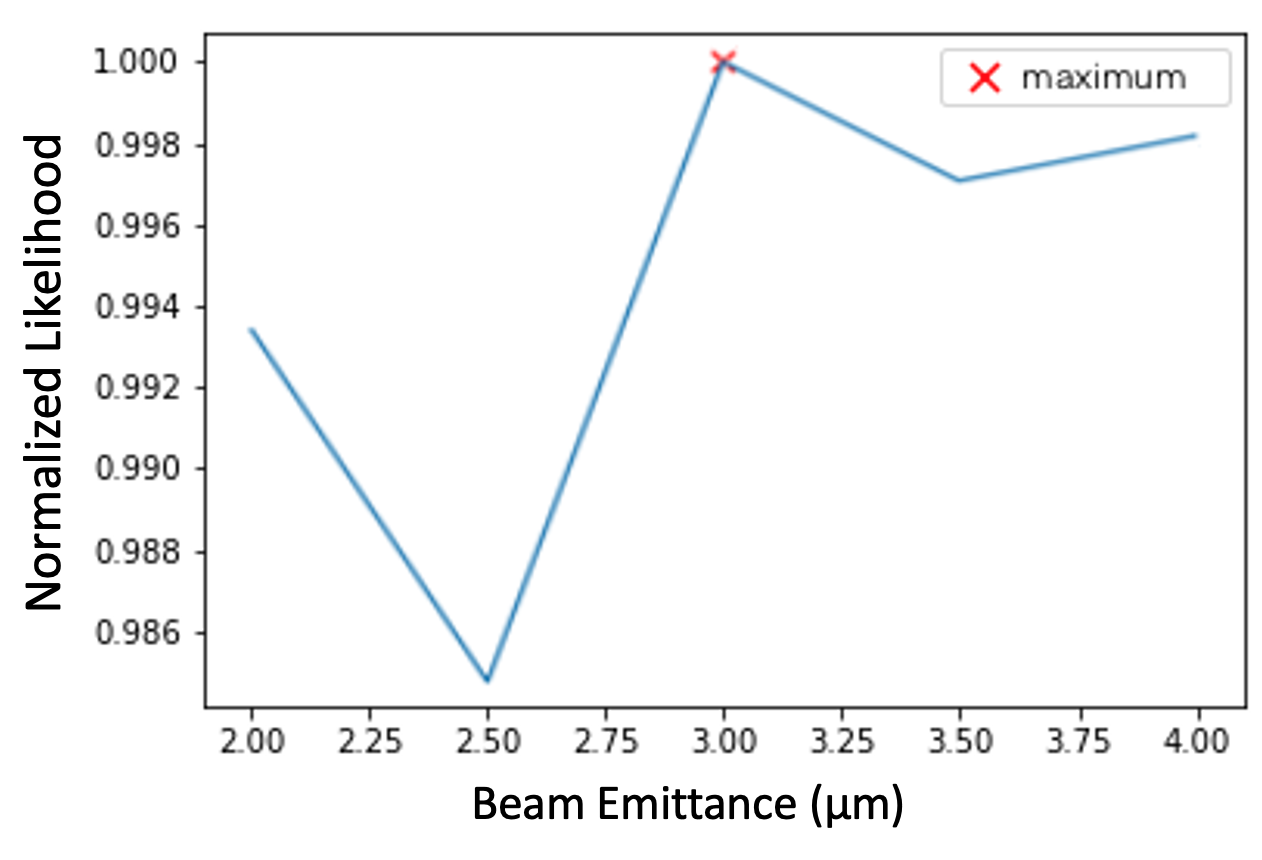}
    \caption{The normalized likelihood function used to identify a test beam emittance of 3 $\mu m$ from 1D radiation spectra.}
    \label{fig:emit_mle}
\end{figure}

In the case of emittance predictions, the spot size of the beam was varied along with the emittance in order to maintain a constant beta-function, which is given by $\beta=\frac{\sigma^2}{\epsilon}$ \cite{betafunction}. That is, for every $\epsilon_\mathrm{new}$,

\begin{equation}
  \sigma_\mathrm{new}=\sqrt{\frac{\sigma_0^2\epsilon_\mathrm{new}}{\epsilon_0}}
\end{equation}

where $\sigma_0$ and $\epsilon_0$ are constant. The results of this MLE, with a test emittance of 3 $\mu$m (the emittance for the matched beam), are shown in Fig. \ref{fig:emit_mle}. Unsurprisingly, the MLE successfully identifies emittance in this basic case. This somewhat trivial case changes only the relative error encountered compared to the beam size analysis - the emittance error is larger by a factor of two compared to the spot size. In less trivial cases, where the beam is mismatched to the focusing beta-function, this approach will be more interesting. 

Having demonstrated that MLE can correctly identify different beam parameters with different types of betatron radiation data, the next step is to develop the method to provide more precise parameter estimations. The likelihood functions shown in Fig. \ref{fig:2d_mle} could only predict the nearest tenth of a $\mu m$. One possible method of doing this would be using the Nelder-Mead method, which starts with an initial guess for the parameter value, then attempts to converge to the actual value. This method involves running simulations for each iterative guess and comparing the results to the test spectrum until arriving at a final value. This method would theoretically allow for more precise estimation in fewer simulations, compared to performing MLE with a grid sampling of parameter value guesses that aims to achieve the same precision.

However, the attempt to implement this algorithm ultimately failed to accurately predict parameter values because it was too easily affected by noise in the likelihood function. Furthermore, like gradient descent methods, the Nelder-Mead algorithm has difficulty finding the global maximum in a function with many local maxima. The likelihood functions this work deals with indeed contain many local maxima due to randomness in the simulations. Therefore, increased precision in parameter estimations is in the end needed by simply accumulating a large amount of simulation data for various parameter values.  

\begin{figure}
    \centering
    \includegraphics[width=\columnwidth]{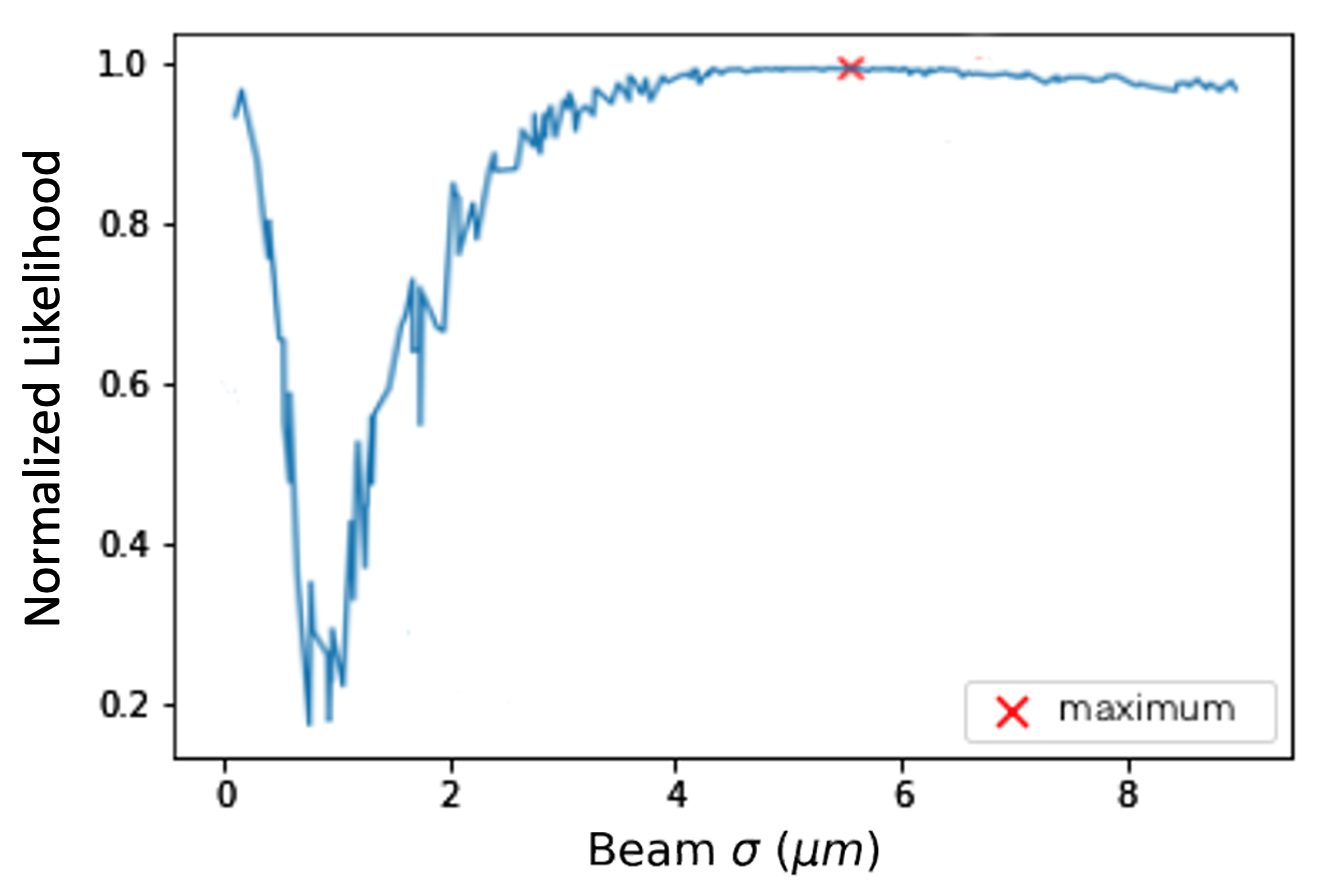}
    \caption{The normalized likelihood function for identifying an test beam spot size of 5.464 $\si{\mu m}$ using 1D radiation spectra. The MLE algorithm predicts a spot size of 5.556 $\si{\mu m}$, using 310 sets of reference data.}
    \label{fig:mle_new_ex}
\end{figure}

\begin{figure}
    \centering
    \includegraphics[width=\columnwidth]{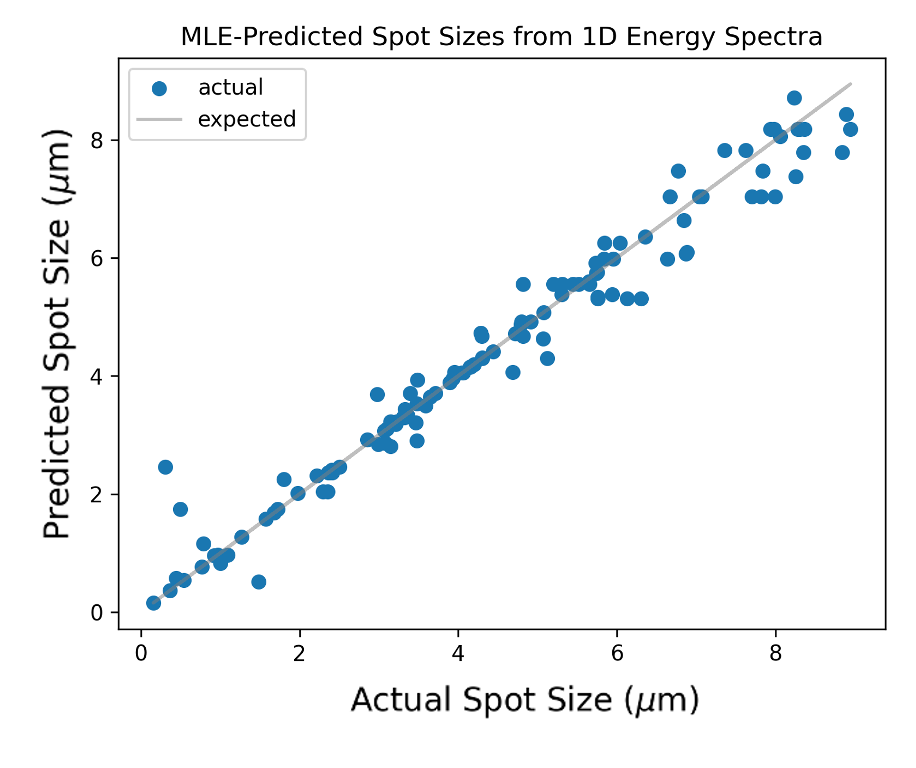}
    \caption{Overall spot size prediction results of the MLE algorithm used with 310 sets of reference data and 120 test cases in the form of 1D radiation spectra. The MSE for these predictions is 0.186 $\si{\mu m^2}$.}
    \label{fig:mle_big}
\end{figure}

In order to obtain increased precision in predicting spot size as described by the method above, a total of 310 simulations are run with only the spot size varied between them. The spot sizes were chosen randomly in a uniform distribution between the values 0.1 $\si{\mu m}$ and 9.0 $\si{\mu m}$. Together, the training spectra obtained from these simulations can be used to calculate a likelihood function that on average can predict to within 0.026 $\si{\mu m}$. In addition, 120 different ``test'' simulations were run, with spot sizes chosen similarly. The results of the likelihood plot comparing all 310 initial simulations to one of these test simulations, which simulates a beam with spot size 5.464 $\si{\mu m}$, is shown in Fig. \ref{fig:mle_new_ex}. 

Here, the MLE examines one-dimensional radiation spectra, and the MLE prediction on spot size is accurate, converging to 5.556 $\si{\mu m}$, or within the resolution margin. The overall results of using the 310 initial simulations to predict the spot sizes of all 120 ``test'' simulations by performing MLE with their 1D radiation spectrum data is displayed in Fig. \ref{fig:mle_big}. The \textit{expected} line in this figure represents perfect predictions, where the predicted spot sizes exactly match the actual spot size. The overall mean-squared error (MSE) for these 120 predictions is 0.186 $\si{\mu m^2}$, and the predictions appear relatively accurate, except in the region below 1 $\si{\mu m}$. Here a few predictions are significantly different than the actual spot sizes. It is noted in this regardthat it becomes physically difficult to distinguish between small spot sizes below a certain point due to the minor differences found by varying $K$ in small $K$ spectra. This is an issue that recurs in later analyses. 

\section{Reconstruction of Beam Parameters Using Machine Learning}
\label{sec:ML}

\begin{figure*}[t]
    \centering
    \includegraphics[width=\textwidth]{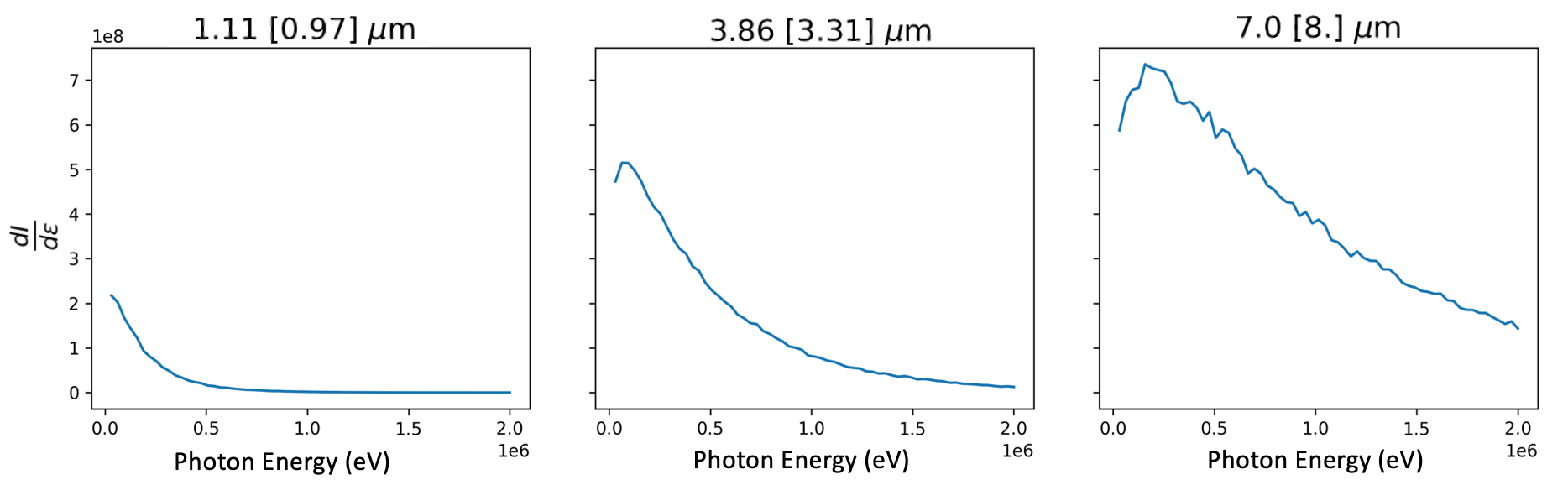} \\
    \includegraphics[width=\textwidth]{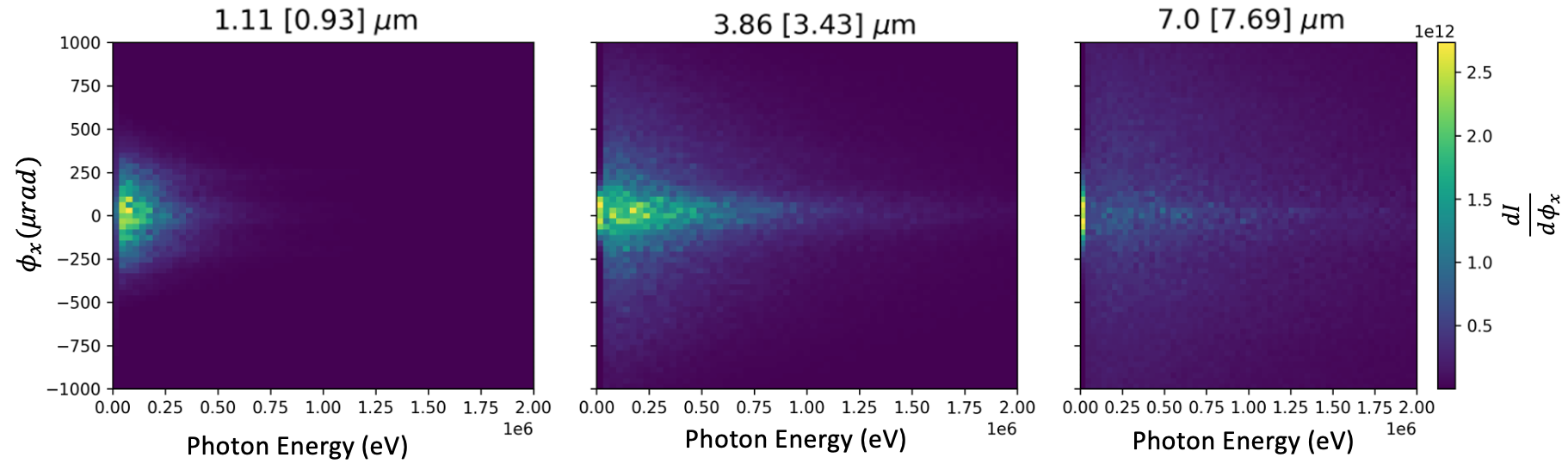}
    \caption{Examples of test 1D and double differential spectra for beams of varying spot sizes are shown. The spectra are each labeled with their true spot size and spot size as predicted by either the direct 1D ML model or the direct double differential ML model. For each plot, the predicted spot size is shown in brackets to the right of the actual spot size.}
    \label{fig:test_sample}
\end{figure*}

While the MLE method of beam parameter reconstruction identified several beam parameters from different types of betatron radiation data, this method is limited in its prediction ability. This method can only predict parameter values for which simulation data is already provided as a reference. Therefore, machine learning is also explored as another method of extracting beam parameters from different plots of betatron radiation.

The machine learning tasks use densely-connected neural networks, which are comprised of several layers containing any number of nodes. In a densely-connected neural network, each node is immediately connected to every node in the layers. These neural networks are built using the open-source machine learning platform TensorFlow \cite{tensorflow2015-whitepaper}. The exact architecture of the fully connected network was determined by selecting the best-performing model with different configurations. The free parameters of the models are the number of layers, the size of each layer, the activation function of each layer, whether or not to add bias to the layer, the learning rate of the model, the maximum output value we normalize to, and we normalize to the maximum data value. 

We trained all the different models with these parameter sets and selected the best-performing model based on the final MSE. The neural networks chosen use rectified linear unit, or ReLU, activation functions. An activation function determines the relationship between the input and output of each node. In particular, the ReLU activation function a(x) is defined as: 
 
\begin{equation}
   a(x) =    
        \begin{cases}
            0 & x\leq 0 \\
            x & 0 < x \\
        \end{cases}
\end{equation}

For training the models, the neural networks use the Adam optimizer with a learning rate of 0.001. The neural networks also use MSE to calculate the loss, or cost, function, which the learning process aims to minimize.

The radiation data in the form of double-differential and 1D spectra were used to train and test two separate models. Similar to the MLE algorithm, these models use the raw intensity values from the spectra as direct inputs to the model, as opposed to, for example, images of the spectra as inputs, which is explored later. Therefore, the 1D and double differential ML models discussed here will be referred to as the direct 1D ML model and direct double differential ML model, respectively. The direct 1D model had three hidden layers (layers between the input and output layers) of 460, 246, and 128 nodes, while the double differential case model had two hidden layers of 800 and 100 nodes. Radiation data was generated using the custom tracking code for 120 different spot sizes ranging from 0.1 $\si{\mu m}$ to 9.0 $\si{\mu m}$. 

Of the 120 sets of data generated, 90 were randomly designated as training cases, and 30 were randomly designated as test cases. A 1D and a 2D double-differential spectrum were extracted from the simulation data for each spot size. The trained model was then used to predict unknown spot sizes for the test data sets. A random sample of the 1D spectrum and double-differential spectrum test cases, along with the actual and predicted spot sizes for each test case, is shown in Fig. \ref{fig:test_sample}. 

\begin{figure}[h!]
    \centering
    \subfloat[Direct 1D ML Model]{\includegraphics[width=\linewidth]{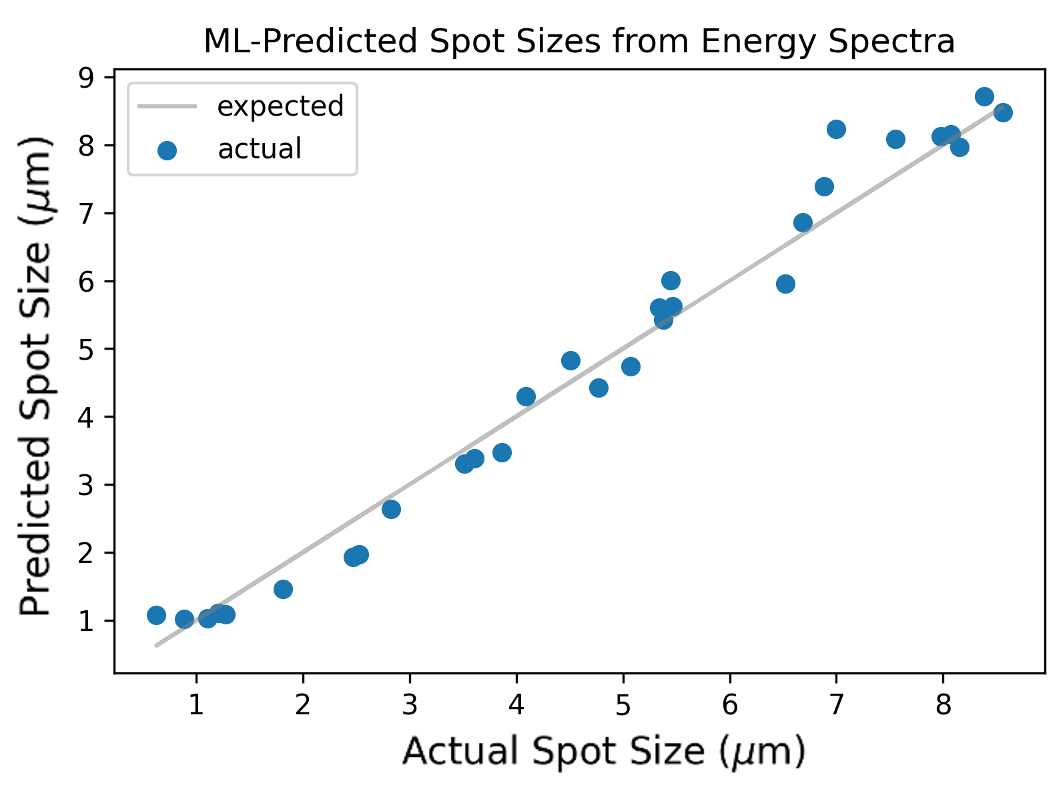}} \\
    \subfloat[Direct Double Differential ML Model]{\includegraphics[width=\linewidth]{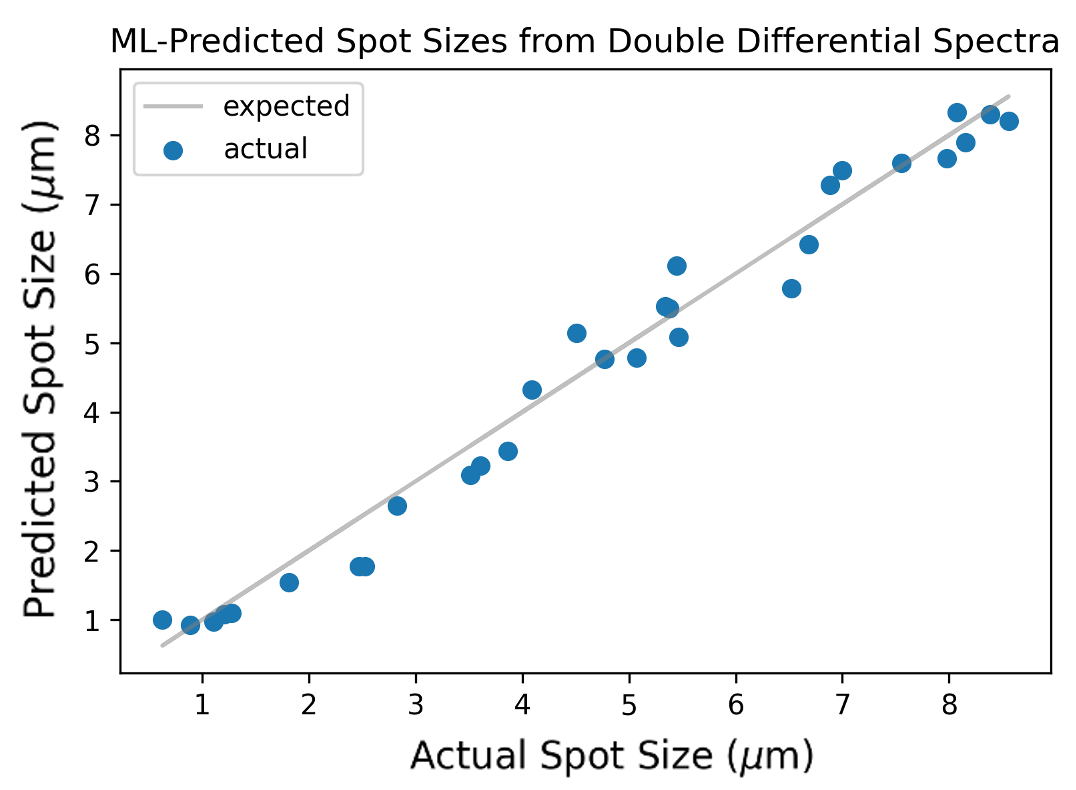}}
    \caption{ML model prediction results for spot size using 90 training cases and 30 test cases across two plot types. The direct 1D ML model was able to predict the spot sizes with an MSE of 0.1556, while the direct double differential ML model was able to predict spot sizes with an MSE of 0.1482.}
    \label{fig:actual_predict}
\end{figure}

The prediction results for all 30 test cases are then represented in Fig. \ref{fig:actual_predict}, which shows the model's accuracy in the case of both the 1D and DDS plots. A similar plateauing is seen in the prediction plots for both models at small spot sizes, under 2 $\si{\mu m}$. In addition, although the spot size predictions in Fig. \ref{fig:test_sample} (shown in square brackets above each plot) differ between the two models, consistently varying from the actual spot size in the same direction. Figure \ref{fig:actual_predict} also suggests that the two models gave similar predictions for most test cases, which implies that the additional information provided by the 2D DDS does not in this simple case significantly improve ML predictions. This would make the 1D spectrum a more natural choice for a first-pass analysis for ML model predictions, because it provides similar results while employing fewer data and less computation.

\begin{figure}
    \centering
    \includegraphics[width=\columnwidth]{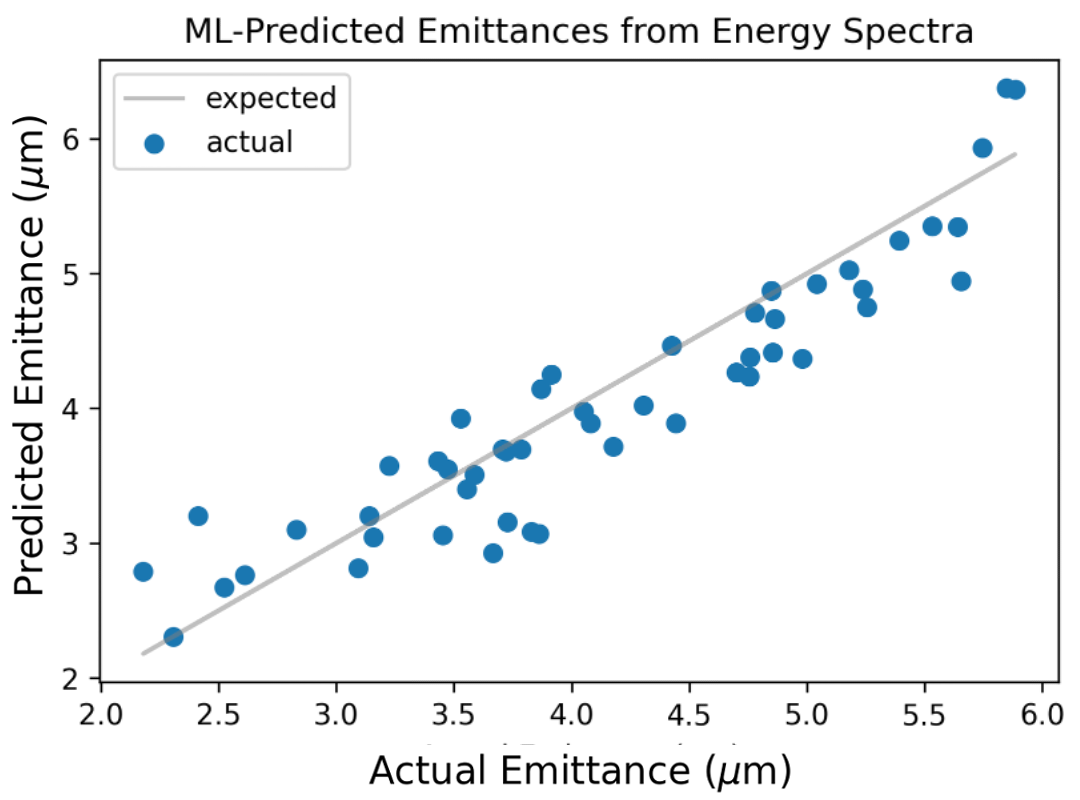}
    \caption{ML model prediction results for emittance using 1D radiation spectrum with 130 sets of training data and 50 test cases. The mean-squared error of the predictions is 0.151 $\si{\mu m^2}$.}
    \label{fig:ml_emit}
\end{figure}

Similar ML models were also trained using the same neural network structure to predict values for emittance. Values for emittance were randomly generated between 2.0 $\si{\mu m}$ and 6.0 $\si{\mu m}$ and used to produce sets of simulation data for the ML prediction method. Figure \ref{fig:ml_emit} shows the prediction results for 50 test cases using a model developed from 130 sets of 1D radiation spectrum training data. The mean-squared error across the 50 test cases was 0.151 $\si{\mu m^2}$. 

\begin{figure}
    \centering
    \includegraphics[width=\columnwidth]{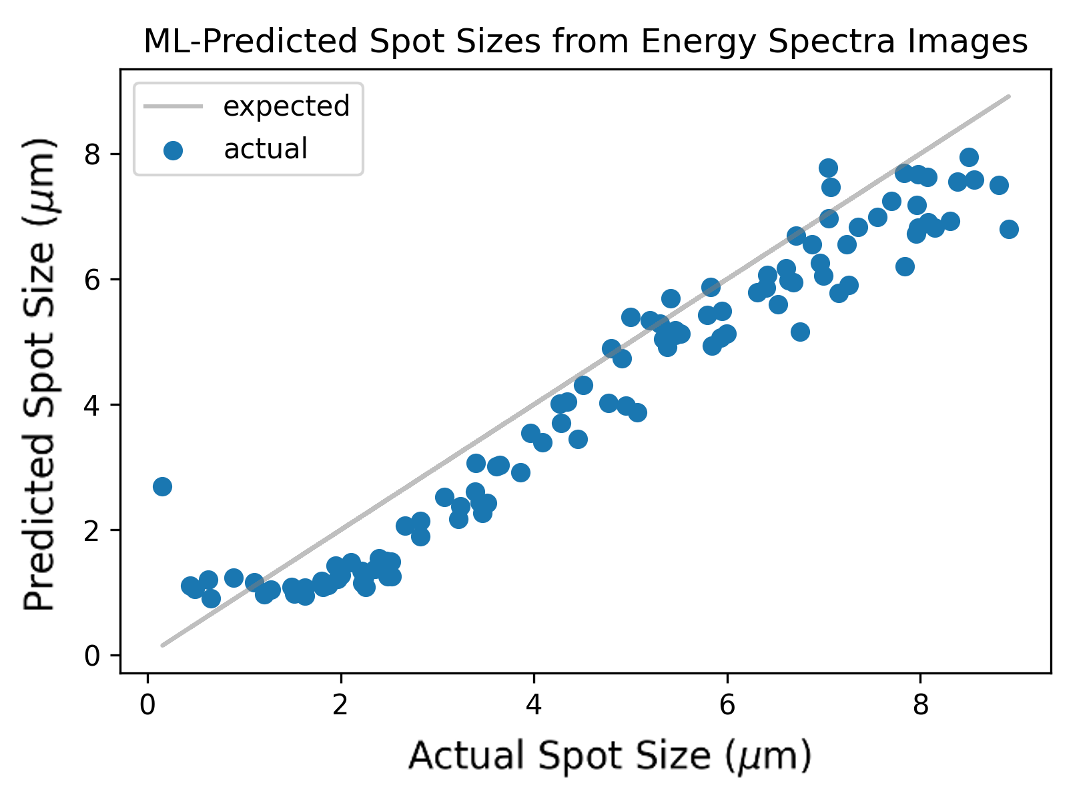}
    \caption{150 $\times$ 150 image input ML model prediction results for spot size. The results are similar to predictions from raw spectrum data, such as those displayed in Fig. \ref{fig:actual_predict}(a). The MSE of the predictions is 1.63 $\mu m^2$.}
    \label{fig:image_ml}
\end{figure}

In addition to the direct ML models, another model was trained to predict spot sizes from 150 $\times$ 150-pixel images of the radiation spectra rather than from the spectrum data itself, which allows for broader application of these ML beam diagnostics. For this 150 $\times$ 150 image input model, the neural network contains six hidden layers with 200, 100, 100, 20, 20, and 10 nodes, and the model was trained over 100 epochs using the loss function in Eq. \ref{loss_func_eq}.

\begin{equation}
    \label{loss_func_eq}
    L(y, \hat{y}) = \frac{1}{N} \left[\sum_{n=1}^{N} (y_i - \hat{y}_i)^2 \right] + \lambda\sum_{j=1}^n|w_j|^2
\end{equation}

The first term represents the standard MSE loss function, while the second represents L2 regularization, which aims to counter over-fitting. Over-fitting occurs when the model fits too exactly against the training data and generalizes poorly to the test data \cite{overfitting_ibm}. L2 regularization, also known as ridge regression, adds a penalty of $\lambda\sum_{j=1}^n|w_j|^2$ to the neural network's loss function. Where \textit{n} is the number of nodes, $w$ is the weight of each node, and $\lambda$ is the L2 regularization factor, which in this case has a value of 0.01 \cite{van2017l2}. By introducing a squared weight term to the loss function, L2 regularization decreases the weight of nodes and effectively simplifies the model so that it does not overfit the training data. 

Before being input into this model, the images were converted to a grayscale, and the value for each pixel was normalized to a value between 0 and 1. The prediction results for this image-based ML model are displayed in Fig. \ref{fig:image_ml}, and the MSE of the predictions is 1.63 $\mu m^2$. The previously observed inaccuracy at small spot sizes continues to be observed in this model.

The results for both MLE and ML predictions displayed in Fig. \ref{fig:mle_big} and Fig. \ref{fig:image_ml} show a tail below $\sim1\ \mu m$, where the predictions all tend to be higher than the actual spot size values. As mentioned above, the persisting inaccuracy at low spot sizes is likely related to the condition $K<<1$, For the purpose of parameterizing the spectrum, here we may take $K_\sigma=\gamma k_\beta \sigma$, as $\sigma$ is the most probable value for the betatron amplitude. 

\begin{figure}
    \centering
    \includegraphics[width=\columnwidth]{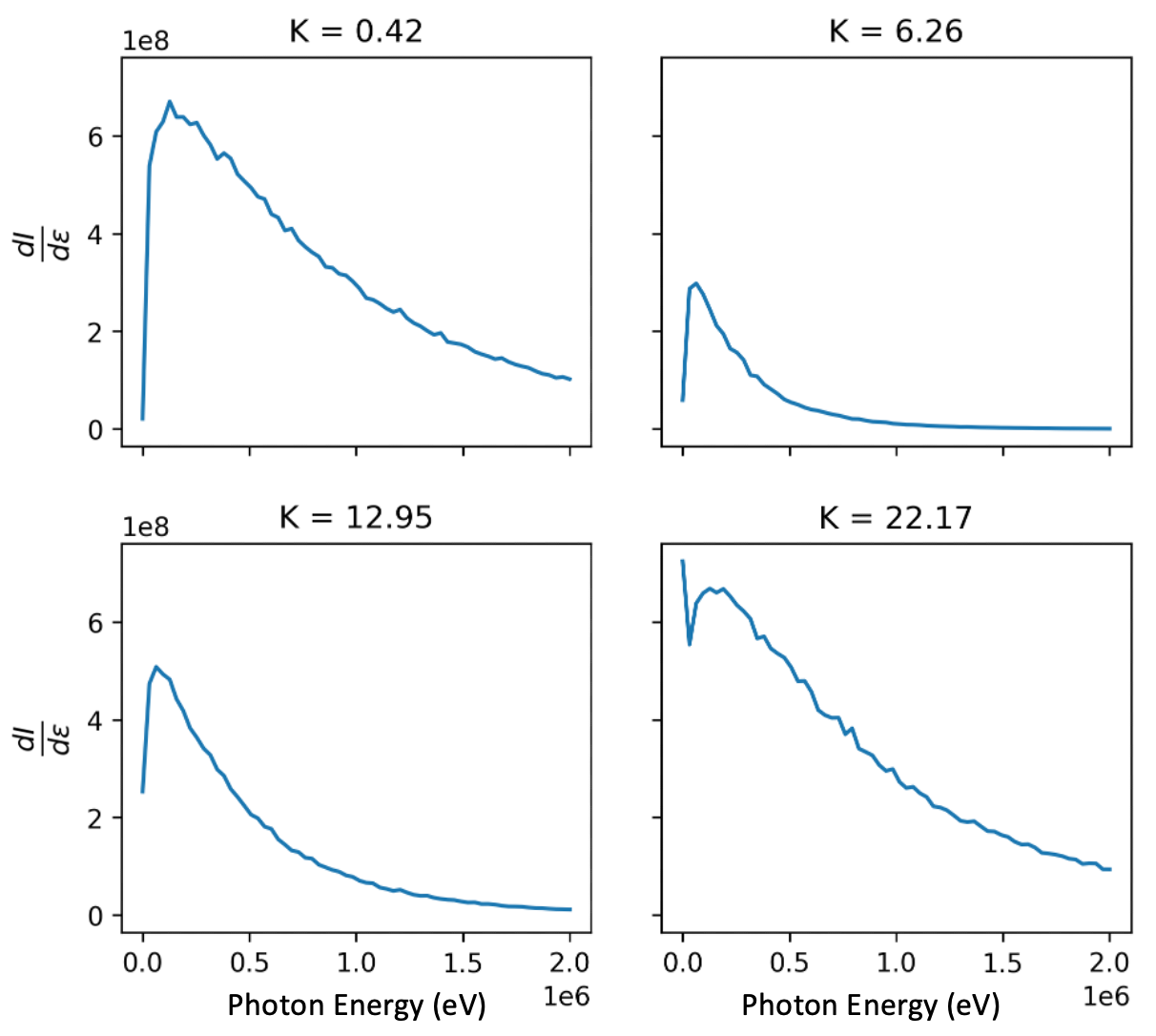}
    \caption{Examples of K values, as calculated by Eq. \ref{k_sig_eq} from each spectrum's simulation parameters, for four different radiation spectra.}
    \label{fig:k_values}
\end{figure}

\begin{figure}
    \centering
    \includegraphics[width=\columnwidth]{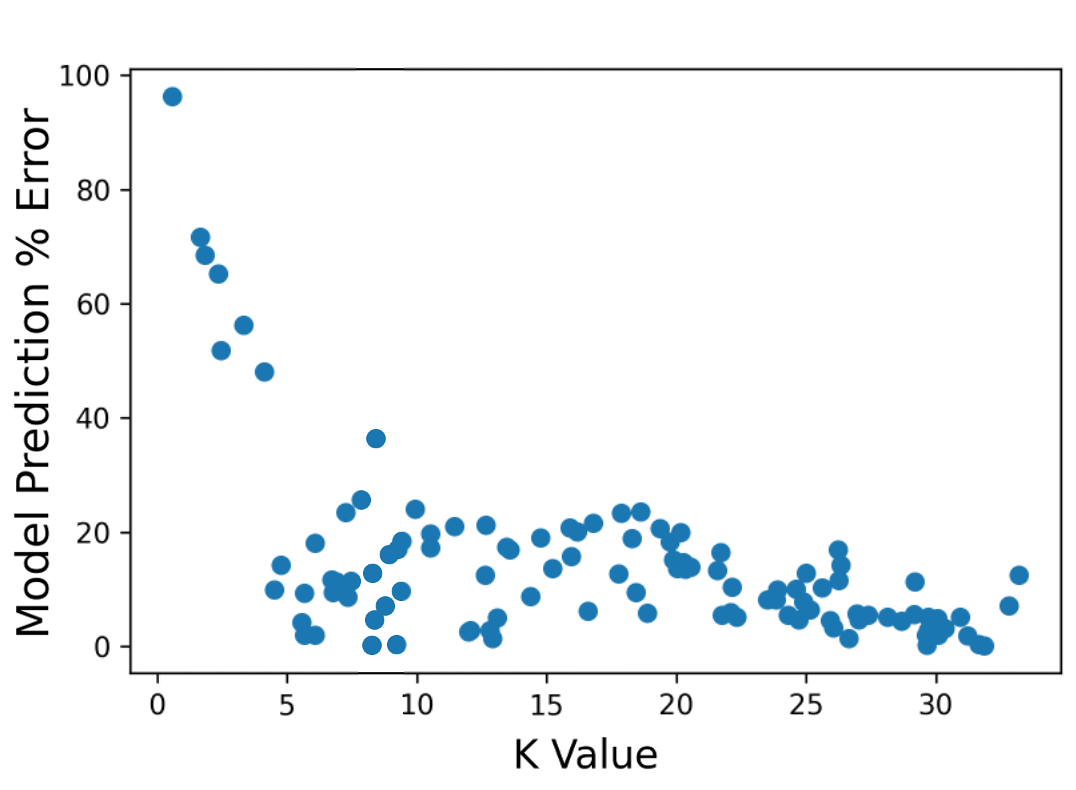}
    \caption{Error in the image input ML model predictions for 1D radiation spectra with different K values. The ML model predicts much worse at lower K values.}
    \label{fig:k_vs_error}
\end{figure}

In the PWFA blowout regime, the betatron oscillation amplitude-dependent \textit{$K_{\sigma}$} values can be therefore calculated as follows:
\begin{equation}
\label{k_sig_eq}
    K_\sigma=\sqrt{\frac{E_be^2n_0}{2\epsilon_0}}\frac{\sigma}{m_{e}c^2}
\end{equation}
The simulated beams' variations in $\sigma$ determined their variations in $K_\sigma$. Figure \ref{fig:k_values} shows examples of radiation spectra and their associated $K_\sigma$ values while Fig. \ref{fig:k_vs_error} shows $K_\sigma$ values for different test cases and the prediction error for each of those cases. The predictions displayed in Fig. \ref{fig:k_vs_error} were made in particular by the 150 $\times$ 150 image input ML model, but the direct 1D ML model predicts similarly. The predictions are pretty inaccurate in the region from K=0 to K=3, and accuracy appears to decrease as $K_\sigma$ increases. For $K_\sigma$ close to unity, the raw $\frac{dI}{dt}$ spectrum of the beam becomes more weakly dependent on $K_\sigma$ and the spot size. This is because, for small $K_\sigma$, the on-axis resonant frequency emitted is proportional to $1+\frac{K^2}{2}$. 

\begin{figure}
    \centering
    \includegraphics[width=\columnwidth]{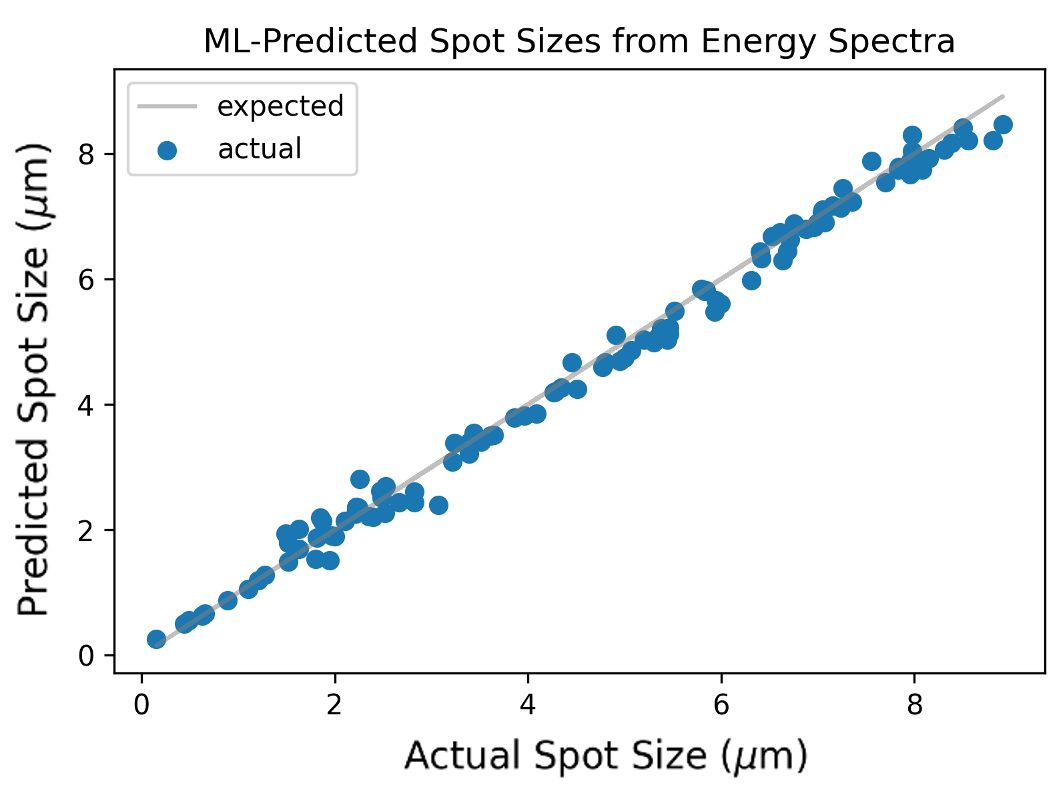}
    \caption{ML model predictions of spot size using 1D radiation spectra normalized according to their K value. The MSE is 0.0513 $\mu m^2$, and the inaccuracies at low spot sizes are no longer present.}
    \label{fig:ml_knorm}
\end{figure}

\begin{figure}
    \centering
    \includegraphics[width=\columnwidth]{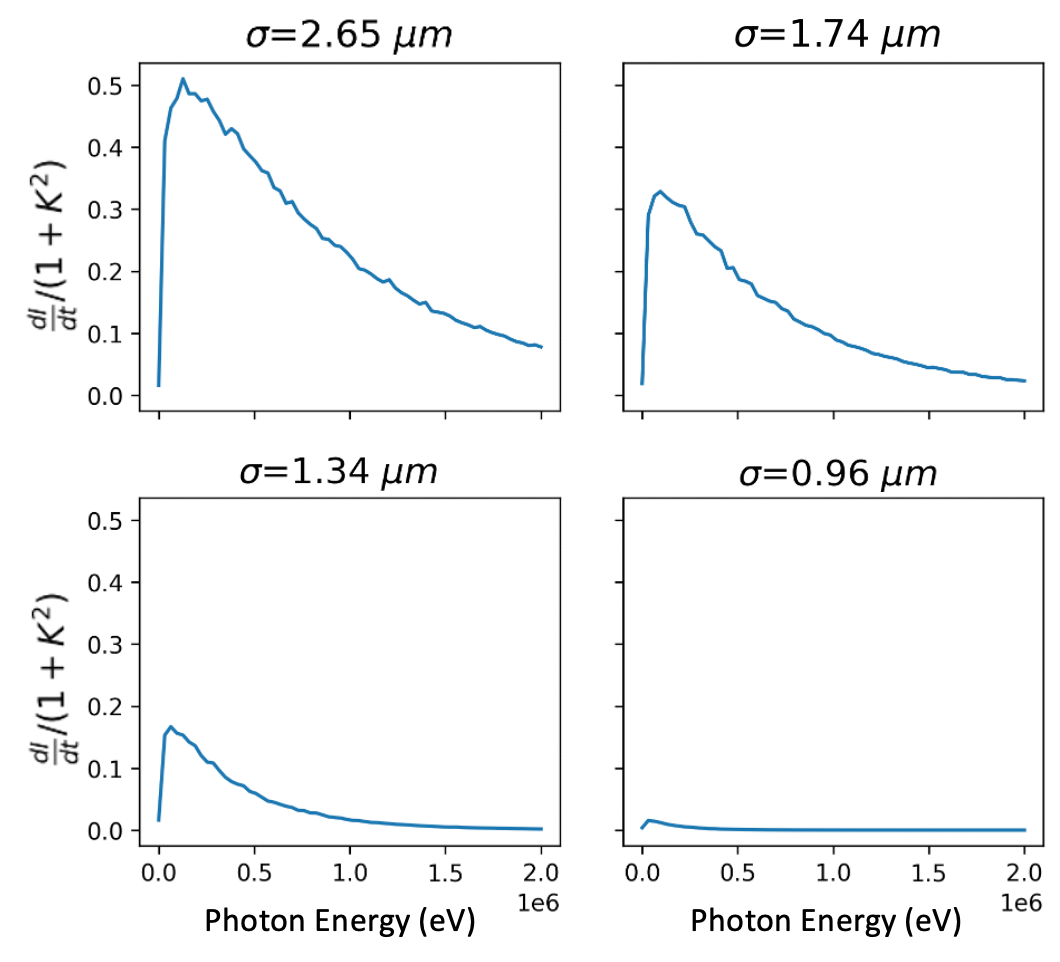}
    \caption{Examples of normalized spectra for different spot sizes. Unlike the raw spectra, the height of the normalized spectra strictly decrease as spot size increases.}
    \label{fig:spec_knorm}
\end{figure}

To improve predictions at low spot sizes, a more sensitive approach at low $K_\sigma$ is needed. As such, the \textit{y}-axis of the $\frac{dI}{dt}$ spectra became $\frac{dI/dt}{1+K_\sigma^2}$ in order to normalize the spectra according to their peak values. These normalized spectra were then used as training and test data for a new ML model. The neural network for the new, normalized spectrum model has four hidden layers with 64, 32, and 10 nodes and was trained over 200 epochs. Like the direct 1D model, it uses the spectrum values as direct inputs into the model. The prediction results for this model are shown in Fig. \ref{fig:ml_knorm}, where the predictions at small spot sizes are no longer significantly larger than the actual spot sizes. Figure \ref{fig:spec_knorm} shows examples of some of these normalized spectra, revealing that, generally, the height of these spectra strictly decreases as spot size increases. 

This is in contrast to the trends observed in the raw spectra, where the heights of the spectra are considerable at tiny sizes, then decrease and increase again as spot size increases. The consistent trend in spectral height makes the spectra much more straightforward for the ML algorithm to differentiate by spot size. The normalization of the spectra according to $K_\sigma$ therefore allows the ML algorithm to successfully differentiate between the emitted spectra of petite spot sizes from each other, eliminating the problem of the inaccurate tail. These MLE and ML beam diagnostic models, based on custom tracker code simulations of Gaussian beams, can thus far only be applied to axis-symmetric beams. They have not been explored yet for perturbed, non-axis-symmetric beams.  Application of similar diagnostic methods to asymmetric beams will be investigated in future work.

These successes with the ML models in predicting spot sizes through limited information demonstrated their reliability in predicting relationships between essential data sets. In the cases of new spectrometers built for FACET-II (the Compton and pair production spectrometers), ML models were considered to estimate the nature of the relationship between incoming electron and gamma energies and the spectrometer responses. Next, in Sec. \ref{sec:reconstruction} we detail the results for the Compton spectrometer, and in Sec. \ref{sec:Pairspec} the results for the pair spectrometer are given. 

\section{Spectral Reconstruction Applied to FACET-II Compton Spectrometer} \label{sec:reconstruction}

\begin{figure}[h!]
    \centering
    \includegraphics[width=70mm,scale=0.5]{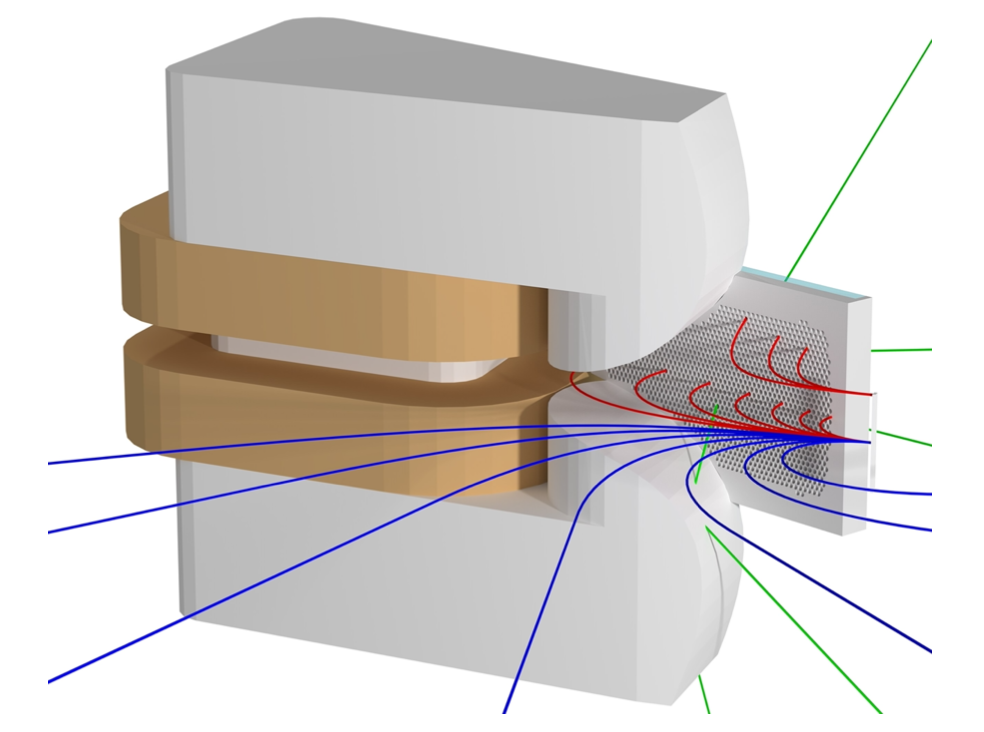}
    \label{fig:uhbemittance}
    \centering
    \includegraphics[width=70mm,scale=0.5]{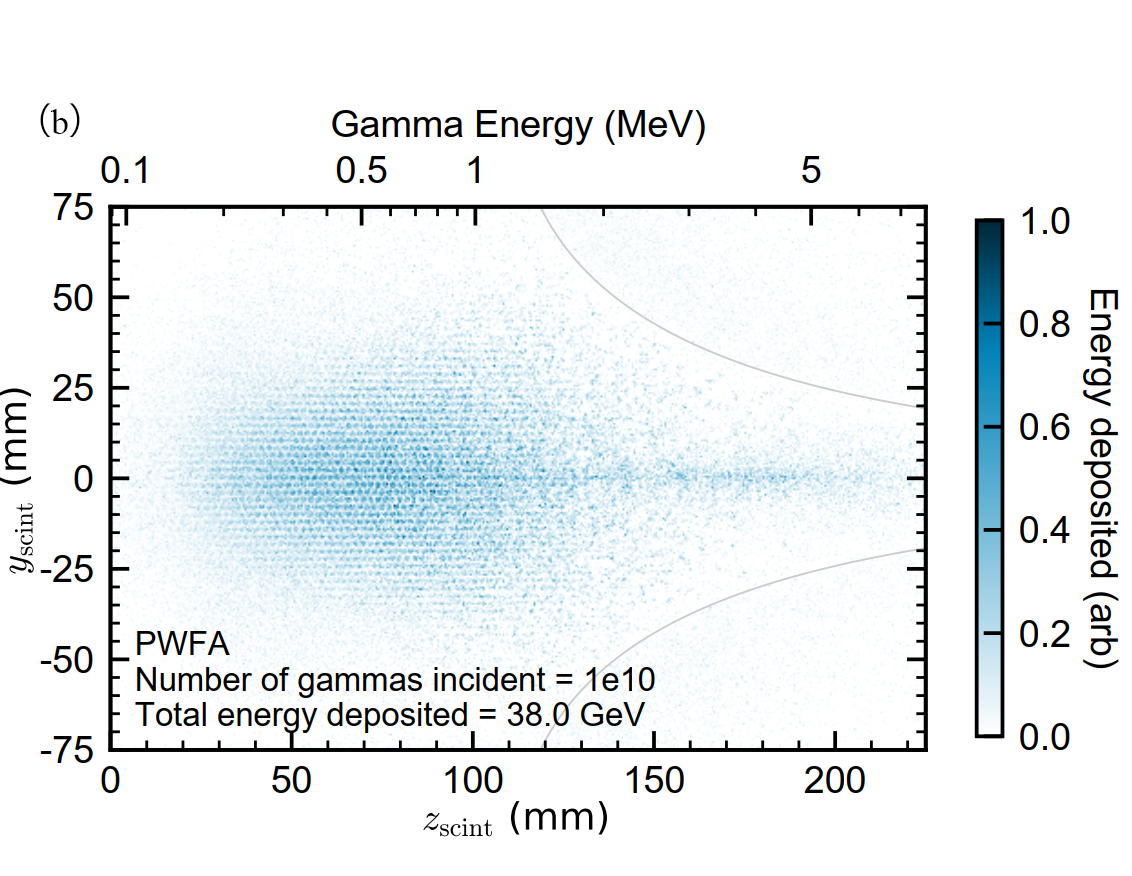}
    \caption{\textbf{(a)} Diagram of UCLA spectrometer. Gammas incident on a beryllium target scatters forward Compton electrons (red), which are bent in a sextupole field and collimated at the focal plane, where a scintillator is located. \textbf{(b)} Simulated image on the Compton scintillator produced by a Geant4 model of the spectrometer.
}
    \label{fig:spectrometer}
\end{figure}

A Compton spectrometer has been developed at UCLA for current implementation at FACET-II \cite{naranjocompton}. To evaluate its expected utility, this work presents two approaches for reconstructing the double differential spectrum from the energy deposition pattern provided by the spectrometer. In the Compton spectrometer, as noted, a sextupolar magnetic field is used to bend the Compton electrons from the converter, giving a compact system that permits a broad range spectral analysis of charged particles. This is shown schematically in Fig. \ref{fig:spectrometer}.  The Compton spectrometer is the first component of a modular magnetic geometry that can be utilized first in lower energy ($<30$ MeV, Compton range) measurements. It can then be extended to the GeV range by adding a downstream spectrometer magnet that permits pair-based measurements. This is a natural separation of components. In the sub-30 MeV energy range, the conversion target employed predominately produces Compton electrons from the primary photon spectrum, with the onset of dominant pair-production occurring above 30 MeV. Magnetic analysis of Compton-scattered electrons has been used as early as 1936~\cite{bothe1936nuclear}.  

Over the years, the technique has been deployed in a variety of scenarios of increasing sophistication
\cite{motz1953gamma,ahrens1973compton,morgan1991broad,sale1992wide,corvan2014design,schumaker2014measurements,tan2017conceptual}. 
The current Compton spectrometer is specifically designed to measure double-differential photon spectra over several decades of energy: from hundreds of keV through tens of MeV, with an angular resolution of $100$ ${\mu rad}$. The primary photon radiation emitted from experiments enters the spectrometer from the lower left of Fig. \ref{fig:spectrometer}(a). It interacts with a lithium converter, provoking both Compton scattering and pair production. 

The lower energy photons interact primarily via Compton scattering, and the resulting electrons are bent by the single-sided (electron) Compton magnet into the two scintillators located on the lower left. The Compton spectrometer utilizes the energy and location of gamma-ray-scattered electrons to recover the incident position and energy of the photons and, thus, their DDS energy-angle distribution. At low gamma energies below 1 MeV, the Compton spectroscopy becomes increasingly challenging as the scattering cross-section becomes more isotropic, and electron scattering in the converter target is also of concern. The UCLA Compton spectrometer uses a pixelated collimation system to select only predominantly forward scattered electrons, permitting incident photons as low as 180 keV to be measured. 

An expectation-maximization (EM) based algorithm and a machine-learning-based algorithm have been used to reconstruct the double-differential spectrum from the energy deposition pattern on the scintillators. The reconstruction problem is posed as follows: given $n(d)$, the energy deposited in the finite scintillator bin $d$, and $p(b, d)$, the energy deposition in scintillator bin $d$ of photons coming from energy-angle bin $b$. We want to obtain $\lambda(b)$, the total energy deposited by the gamma photons of each energy-angle bin. In other words, given the overall energy deposition pattern on the scintillators and a basis composed of the deposition pattern for each possible binned energy and angle of incoming photons, we are required to find the true total binned double-differential spectrum of the incoming photons. This is achieved by using two approaches: an EM-based algorithm and a machine learning-based algorithm, both explained below.

\subsection{Expectation maximization for the reconstruction of Compton scattering spectrum}

We have applied the EM-based algorithm to resolving the energy spectrum without angular dependence\cite{vardi1985statistical}. In the Expectation step (E-step), we begin with an initial guess of an emission rate associated with the photon energy bin $b$ to be $\lambda^\mathrm{old}(b)$. Then we can estimate the energy of the photons emitted in energy bin $b$ to be $\hat{n}(b)$. 
\begin{align}
    \hat{n}(b) \sum_{d=1}^D p(b, d) &= \sum_{d}\mathbb{E}[n(b, d)|\lambda^\mathrm{old}(b), n^*(d)]
\end{align}

The sum $\sum_{d=1}^D p(b, d)$ is present because not all photons are collected. Note that $n(b, d)$ are a set of Poisson random variables with sum $n^*(d)$, so each of them follows the binomial distribution with probability $\lambda^\mathrm{old}(b, d)/\sum_{b'}\lambda^\mathrm{old}(b', d)$. We can evaluate the expectation value as
\begin{equation}
    \hat{n}(b) \sum_{d=1}^D p(b, d) = \lambda^\mathrm{old}(b)\sum_d \frac{n^*(d)p(b, d)}{\sum_{b'}\lambda^\mathrm{old}(b')p(b', d)} \label{eq:n_hat}
\end{equation}

In the Maximization step (M-step), if $\hat{n}(b)$ is our estimate for energy emitted in energy bin $b$, it also serves as the estimate for the corresponding emission rate $\lambda(b)$, because we model photon emission as a Poisson process with rate $\lambda(b)$. In other words, when $\lambda(b) = \hat{n}(b)$, it is most likely to observe $\hat{n}(b)$.

To create the complete algorithm, we update $\lambda^\mathrm{new}(b)$ for 50 iterations with the following rule. 
\begin{equation}
    \lambda^\mathrm{new}(b) = \frac{\lambda^\mathrm{old}(b)}{\sum_{d=1}^D p(b, d)}\sum_d \frac{n^*(d)p(b, d)}{\sum_{b'}\lambda^\mathrm{old}(b')p(b', d),}\label{iteration}
\end{equation}
where we substitute $\hat{n}(b)$ with $\lambda^\mathrm{new}(b)$ and rearranged the terms in Eq.~\ref{eq:n_hat}.

In considering different choices for $p(b, d)$, one option for $p(b, d)$ is the energy deposition of mono-energetic gamma-rays, which is the energy deposition in bin $d$ on the scintillator, which comes from the photon energy bin $b$. A Geant4 simulation generates energy deposition.

We also utilized a smoothed basis for $p(b, d)$, i.e., replacing the original delta function in energy bin $b$ with a Gaussian function centered at $b$ with $\sigma = 6$ bins. Under such conditions, $p(b, d)$ is interpreted as the scintillator output of one Gaussian-distributed beam. 

\subsection{Reconstruction of Compton Spectrum Using Machine Learning} 

Machine learning comprises algorithms that capture essential features of an arbitrary function to mimic the learned function's behavior given new inputs. We use here a fully connected neural network to reconstruct both the 1D energy and the 2D energy-position spectra. Here the input for our model is the output of the spectrometer, namely, $n^*(d)$. The algorithm's task is to model the relation between $n*(d)$ and $n(b)$ to generate gamma spectrum $n(b)$ for any given $n*(d)$. 

A fully connected neural network is formed by a series of matrix multiplications ($\textbf{W}_i$) with vectors in the hidden layers ($\text{hid}_i$) plus biases ($b_i$), each attributed to the evaluation of a nonlinear activation function $F$:
\begin{align*}
    \text{Out} &= F(\textbf{W}_1\text{hid}_1+b_1)\\
    &= F(\textbf{W}_1F(\textbf{W}_2\text{hid}_2+b_2)+b_1) = ... 
\end{align*}

The model learns to use the weights $\textbf{W}$ to approximate a general function by adjusting their values towards the minimization of the error in the output.  In generating training data, it is known that only reasonably balanced and sufficiently random training sets may help the model capture the variety of outputs and generalize model capabilities.
\begin{figure}[h!]
    \centering
  {\includegraphics[width=0.47\textwidth]{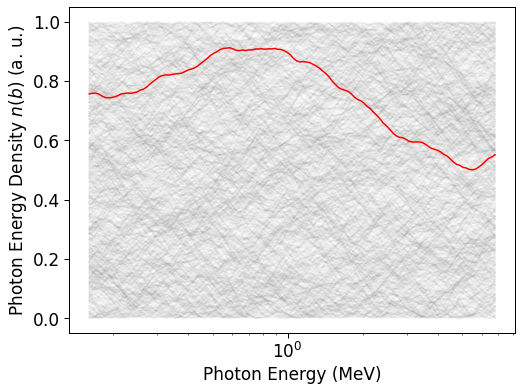}}\\
    \caption{1000 generated spectra used in machine learning and one example are plotted above, in transparent gray and red, respectively. The bounded diffusion simulation ensures that there is no bias towards particular shapes in the curves of training data, as visualized by the evenness of the gray background.}  
    \label{fig:cpt-labels}
\end{figure}

The ground truth for the model output to approximate the energy spectra should be independent of specific shapes to ensure the model's full flexibility. Therefore, the training data set for the one-dimensional energy spectrum-only case was generated by an algorithm that provided uniformly mixed energy spectra as ground truths, as shown in Fig. \ref{fig:cpt-labels}. The corresponding scintillator responses were generated by linearly superposing a data set of mono-energetic responses. This approach is valid because energy deposition is a linear process and a superposition of mono-energetic outputs is thus a valid output from the scintillator. 

To achieve uniform, random, yet reasonably smooth 1D energy spectra, we have employed trajectories of particles diffusing with a drag force in a bounded box. Those trajectories were simulated step-wise with Python using the following velocity update rule: 
\begin{equation*}
v_{i+1} = 0.95 \times (v_i + 0.01 \times (\mathrm{Uniform[0, 1]} - 0.5))
\end{equation*}
Furthermore, we add the elastic reflection at the bottom and top boundary.

After experimenting with different network parameters, the selected model for the 1D (energy only) case has three hidden layers of 128, 128, and 512 nodes, no bias, and ReLU (Rectified Linear Unit) activation.

\begin{figure}[h!]
    \centering
    {\includegraphics[width=0.47\textwidth]{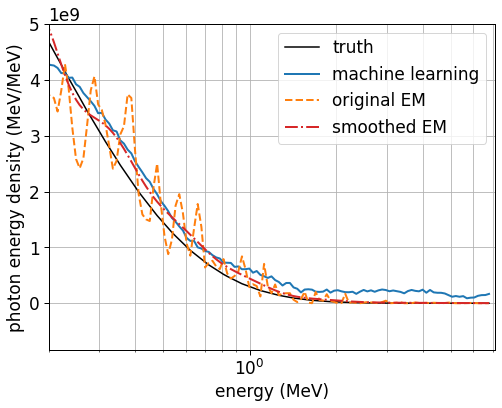}}\\
    \caption{Results of three different reconstruction algorithms for the PWFA experiment. The true PWFA spectrum is plotted in solid black. The blue curve is the predicted spectrum from the ML-based algorithm. The dashed orange curve corresponds to the result of the original EM Algorithm, and the red dash-dot line is the result of the smoothed EM Algorithm. The dash-dot line agrees best with the truth.}  
    \label{fig:cpt-pwfa}
\end{figure}

To illustrate the performance of the above approaches, we must go beyond the simplest models for trajectory generation discussed above. Here we present the reconstruction results for the data set created for a PWFA experiment, including a transversely matched trailing beam, a simulated spectrum for the E300 experiment at FACET-II (Fig. \ref{fig:cpt-pwfa}). We have calculated the point-wise root mean squared (RMS) error between the truth spectrum normalized to have a maximum of 1 and the prediction divided by the same factor. The machine learning result shows an RMSE of 0.055. The RMS error for the original EM algorithm is 0.0878, and that for the smoothed EM algorithm is 0.0354. 

\begin{figure}[h!]
    \centering
   {\includegraphics[width=0.47\textwidth]{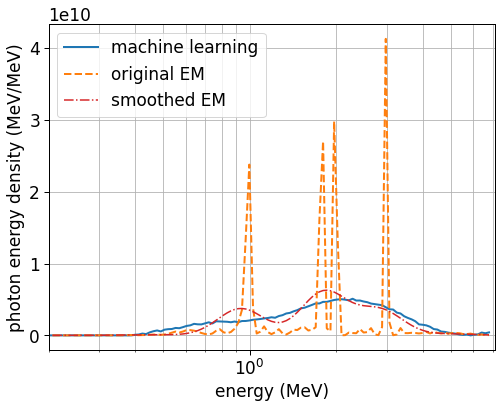}}\\
    \caption{Results of three different reconstruction algorithms for resolving peaks. The blue curve is the predicted spectrum from machine learning. The dashed orange curve corresponds to the original EM Algorithm, and the red dash-dot line results from the smoothed EM Algorithm. The original EM Algorithm is the best at resolving peaks because no smoothing is involved in the process.}  
    \label{fig:cpt-darpa}
\end{figure}

As shown in Fig. \ref{fig:cpt-pwfa}, the smoothed EM algorithm using a Gaussian basis performed the best for reconstructing smooth spectra. However, the original EM algorithm is more capable of resolving discrete peaks, as shown in Fig. \ref{fig:cpt-darpa}, which illustrates a linear inverse Compton scattering spectrum. Therefore, a differing spectral analysis may be needed for each experiment to determine the amount of smoothing appropriate for the reconstruction algorithms.  

\begin{figure}[t]
    \centering
    \includegraphics[width=\columnwidth]{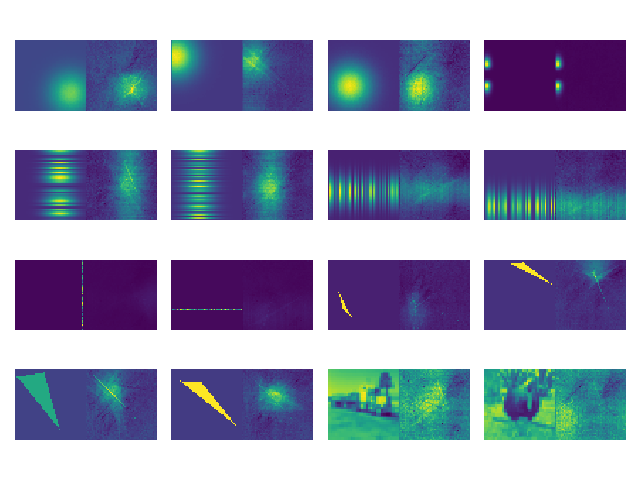}
    \caption{Truth vs. Prediction for 16 different samples in reconstructing double differential spectra via machine learning. For each pair, the truth is on the left square, and the reconstruction is on the right. The last two cases show arbitrarily complex spectra. The model is capable of recovering the general shape of the 2D spectra in many cases, while the more complex cases are more challenging.  }
    \label{fig:cpt-2d}
\end{figure}

Only machine learning has been applied to reconstructing the EEW for an earlier version of the Compton spectrometer, and we present thevpreliminary results obtained in Fig. \ref{fig:cpt-2d}. We have used the total scintillator outputs from Geant4 \cite{tevpermeter}. 

The model can reproduce approximate shapes of the 2D spectra in several cases. 

In its original form with the mono-energetic basis, the EM algorithm performed well with locating peaks but was susceptible to noise and undesirable fluctuations. A smoothed basis reduces the fluctuations and but lessens the ability to discern nearby peaks. On the other hand, machine learning with a fully connected network is a flexible tool for reconstructing different types of spectra. Trained with carefully synthesized data, the model learned to approximate the functioning of the spectrometer. Future work will involve applying the EM algorithm to 2D reconstruction and ML-based algorithms to reconstruct double-differential spectra for an updated spectrometer design. 

\section{Methods: pair spectrometer analysis}
\label{sec:Pairspec}

We now discuss the analyses necessary when using the higher energy pair-production-based spectrometer (PEDRO). It is necessary to approach this problem somewhat differently due to the additional complexity of determining the initial photon energy distribution uniquely. When considering electron-positron pair production from photon interactions with nuclear fields, it is essential to recognize the linear relationship between the energy distribution of photons and the spectrometer's response. This linear relationship can be modeled in Eq: \ref{eq:19} is as follows:

\begin{equation}
    \begin{cases}
      \alpha_{1,1}*x_1 + ... + \alpha_{1,64}*x_{64} = y_1 \\
      \alpha_{2,1}*x_1 + ... + \alpha_{2,64}*x_{64} = y_2 \\
      ... \\
      \alpha_{128,1}*x_1 +  ... + \alpha_{128,64}*x_{64} = y_{128} \\
    \end{cases}
     \label{eq:19}
\end{equation}
where the vector ($x_1$, $x_2$,..., $x_{64}$) represents the gamma energy distribution, $\alpha_{i,j}$ are coefficients, and the elements ($y_1$, $y_2$,..., $y_{128}$) represent the spectrometer's responses. 

In the above system of equations, and throughout the rest of this paper, the \textit{x}-vector will refer to the original distribution of photons based on their energies. While the energy values will become more relevant when interpreting the model's output, the numerical values that split the energy distribution into logarithmic bins were chosen for convenience and did not impact the model's logic or construction. The \textit{y}-vector will always refer to the electron-positron spectrum that PEDRO outputs in response to the corresponding \textit{x}-vector, or incoming photon energy distribution. 

Additionally, this linear relation permits building a matrix that can be used to generate any number of training and test cases for training and assessing the accuracy of different models when recovering the photon energy distributions from spectral responses. The following matrix with coefficients $\alpha_{i,j}$ will be referred to as the response matrix $R$ since it models how the spectrometer will respond to a given photon energy distribution. 

\begin{equation} \label{eq:pedro}
    \begin{bmatrix} 
        \alpha_{1,1} & \alpha_{1,1} & \cdots & \alpha_{1,64} \\
        \alpha_{2,1} & \alpha_{2,1} & \cdots & \alpha_{2,64} \\
        \vdots       & \vdots        & \ddots &  \vdots \\
        \alpha_{128,1} & \alpha_{128,1} & \cdots & \alpha_{128,64} \\
    \end{bmatrix}
    \times
    \left[
        \begin{array} {c} 
            x_1 \\ 
            x_2 \\ 
            \vdots \\ 
            x_{64} 
        \end{array}
    \right] 
    = 
    \left[ 
        \begin{array}{c}
            y_1 \\ 
            y_2 \\ 
            \vdots \\ 
            y_{128} 
        \end{array} 
    \right]
\end{equation}

To determine the best method of recovering the \textit{x}-vector given a \textit{y}-vector, several computational methods were explored and compared when applied to standard test cases. The three methods included: developing a machine-learning model, combining the model with the MLE algorithm, and applying QR decomposition to the \textit{R} matrix. 

To determine the efficacy of each approach, there were five standardized test cases to which each method was applied. They are listed as follows:

\begin{itemize}
  \item Mono-energetic: The original spectrum contained $10^8$ photons in only the second bin and 0 photons elsewhere. Noise with a level of 100 was added from a Poisson distribution to introduce experimentally relevant variability to the spectrum. 
  \item Arbitrary Case: The original spectrum contained only a random number of photons between 0 and $10^{10}$ in all of the photons. Noise with a level of $10^4$ was added from a Poisson distribution to introduce variability to the spectrum.
  \item Three other smooth cases that were derived from experimental scenarios (nonlinear Compton Scattering, high field quantum electrodynamics, filamentation in plasma). No Poisson noise was added in these cases since they inherently represented noisy data obtained from simulation. 
\end{itemize}

\noindent The frequency for the mono-energetic case was chosen at random. 

For each of these test cases, and in future experiments, the spectrometer's ability to reconstruct frequencies that were within a factor 1e04 from the spectrum's peak was considered. In any given spectra, that range alone can be reliably reconstructed without the interference of upstream noise. This lower readout limit is indicated in each of the following graphs by a red dashed line, and any spectrometer behavior of spectra below that line was not used to determine the reconstruction method's efficacy.

\subsection{Machine learning approach for PEDRO analysis}

Using Eq.~\ref{eq:pedro}, training data was synthesized by creating an arbitrary energy distribution and multiplying each \textit{x}-vector by the matrix to generate the corresponding spectrum (or \textit{y}-vector) that PEDRO would measure. To simulate real-world noise from scattered electrons during the pair production process, low-level noise vectors (each element assigned a number between 0 and $10^4$ selected from a Poisson distribution) were calculated and added to the \textit{y}-vectors.

Once the training and test data sets were synthesized, the model was then provided the set of \textit{y}-vectors and instructed to calculate which \textit{x}-vector interacted with the spectrometer to give rise to such \textit{y}-vector. 

To determine the optimal settings for the model, including the type of optimizer and number of layers, each model element was varied independently, and the model was trained on the same training data set. Then, the accuracy output from testing the model against the testing data set was used to assess which would produce the most desirable results, with a higher accuracy value indicating the setting was more favorable. 

Table~\ref{table:1} summarizes the model's architecture. The model used the Adam optimizer with a learning rate of 0.005 over 600 epochs. Mean squared error was chosen as the loss function for the model, which was constructed using the Python Keras library. \cite{chollet2015keras}.

\begin{table}[!bht]
	\centering
	\caption{A summary of the ML model to predict incoming gamma spectra based on positron-electron detection}
	
	\begin{tabular}{ c c c  } 
	 \toprule
	 \textbf{Layer} & \textbf{Output} & \textbf{Param} \\ 
	 \textbf{(type, bias, activation)} & \textbf{Shape} & \textbf{Num} \\ 
	 \midrule
	 dense (Dense, true, linear) & (None, 64) & 64 \\ 
	 
	 dense1 (Dense, true, linear) & (None, 64) & 64 \\ 
	\bottomrule
	\end{tabular}
	\label{table:1}
\end{table}

Given the linear relation between the incoming photon energy distributions and the output PEDRO spectra, it was determined that the model should implement only linear activating functions. The number of layers, the output shape, and the number of parameters were all tested through a series of trials. The above configuration construction was determined to have the best predictive power. 

\begin{figure}[t]
  \centering
    \subfloat[Mono-energetic case]{\includegraphics[width=0.47\textwidth]{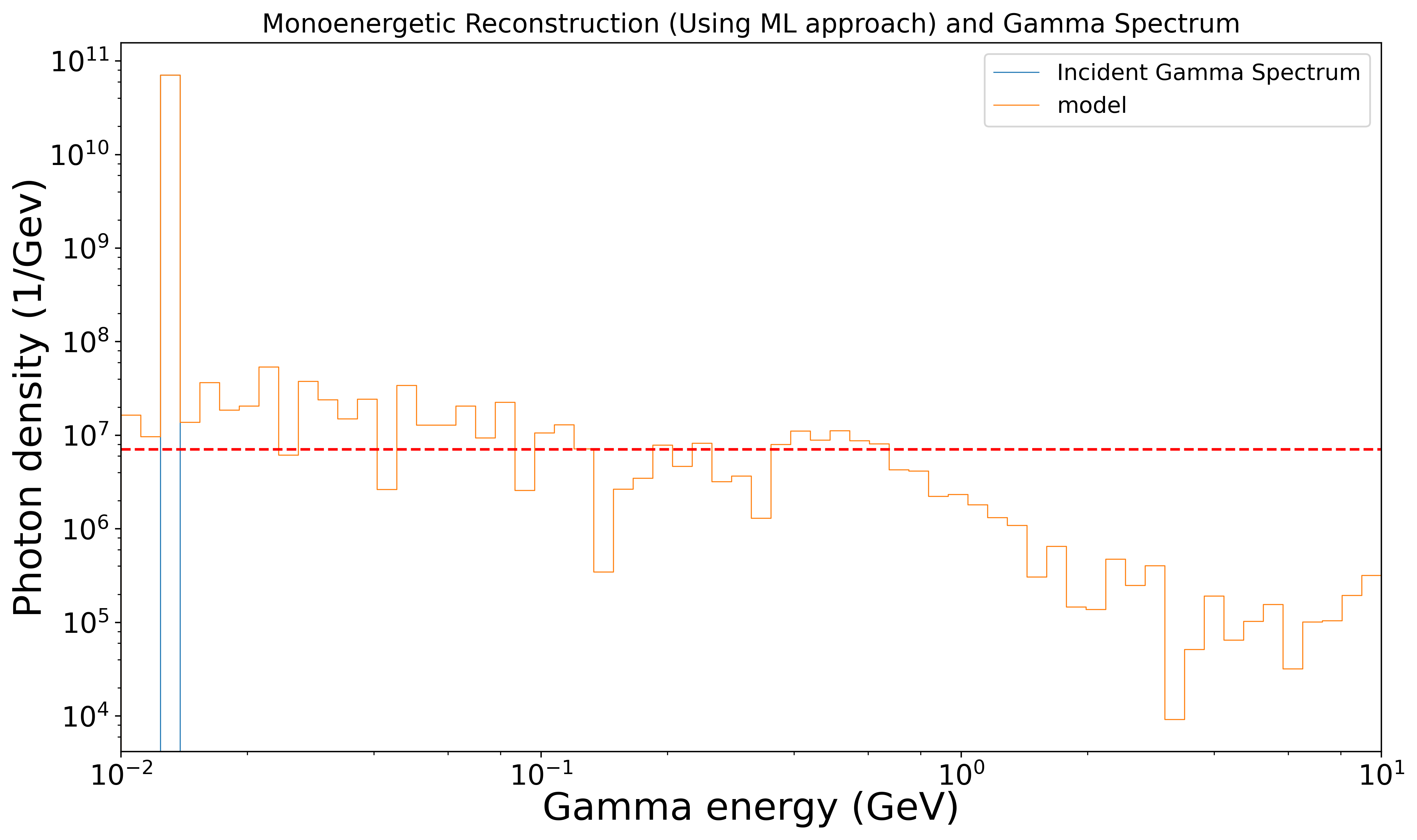}}\\
    \subfloat[Arbitrary Discrete case]{\includegraphics[width=0.47\textwidth]{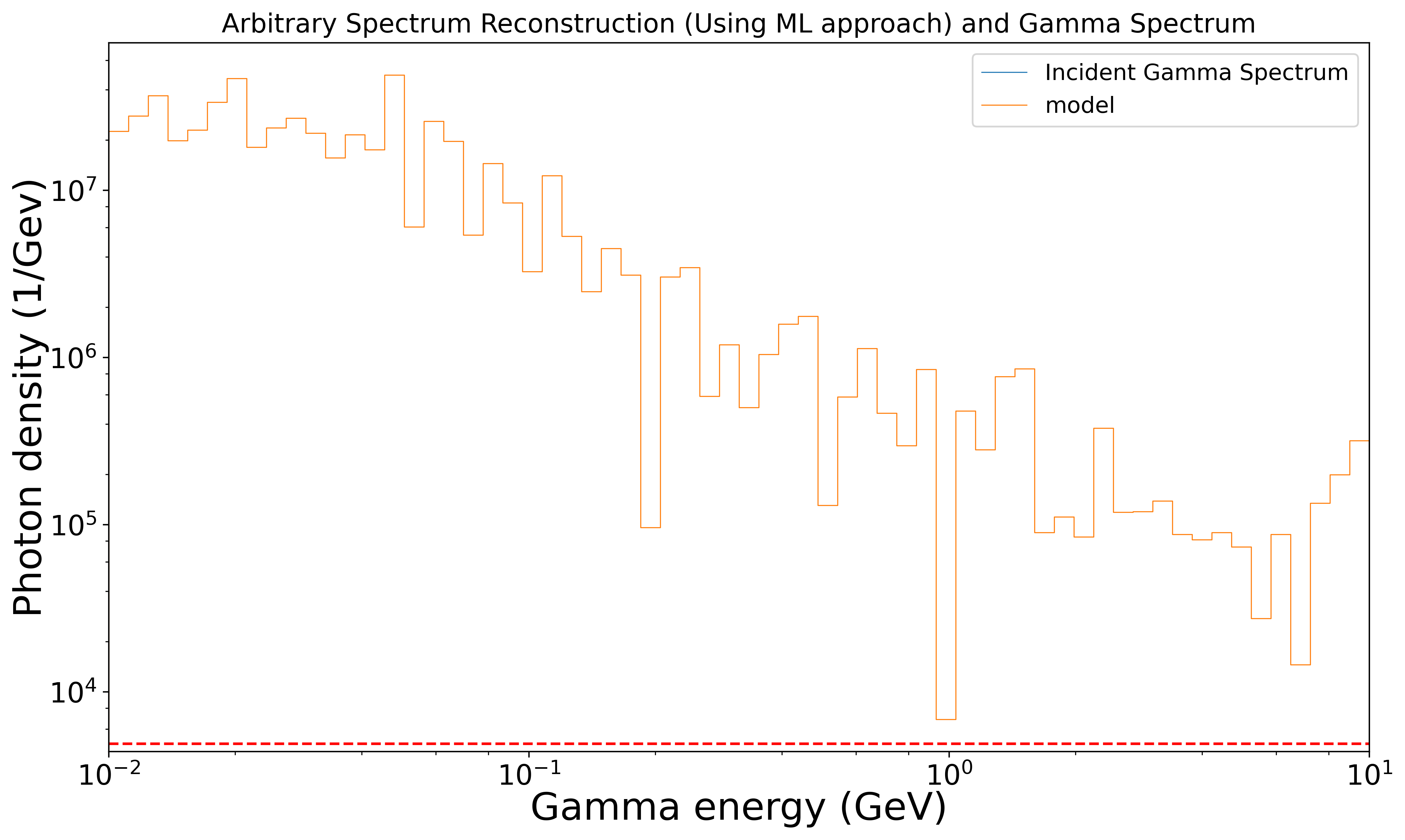}}\\
  \caption{Discrete reconstructed gamma energy distributions using machine learning}
  \label{fig:2}
  \vspace*{-\baselineskip}
\end{figure}

Using machine learning, reconstruction success varies depending on the individual cases. Looking at the discrete cases in Fig. \ref{fig:2}, the trained ML model predicts peaks at the correct bins where the photons were located in terms of their energies in the mono-energetic case. The scale is also accurately reconstructed, followed by predictions of photon frequencies several orders of magnitude lower than the incident peak. In the arbitrary discrete case, the model effectively reconstructs the distribution with only minimal deviations visually apparent in between bins 30 and 40 (approximately corresponding to an energy range 250-670 MeV), as seen in Fig. \ref{fig:2}(b).

\begin{figure}[t]
  \centering
  \subfloat[Non-Linear Compton scattering case]{\includegraphics[width=\columnwidth]{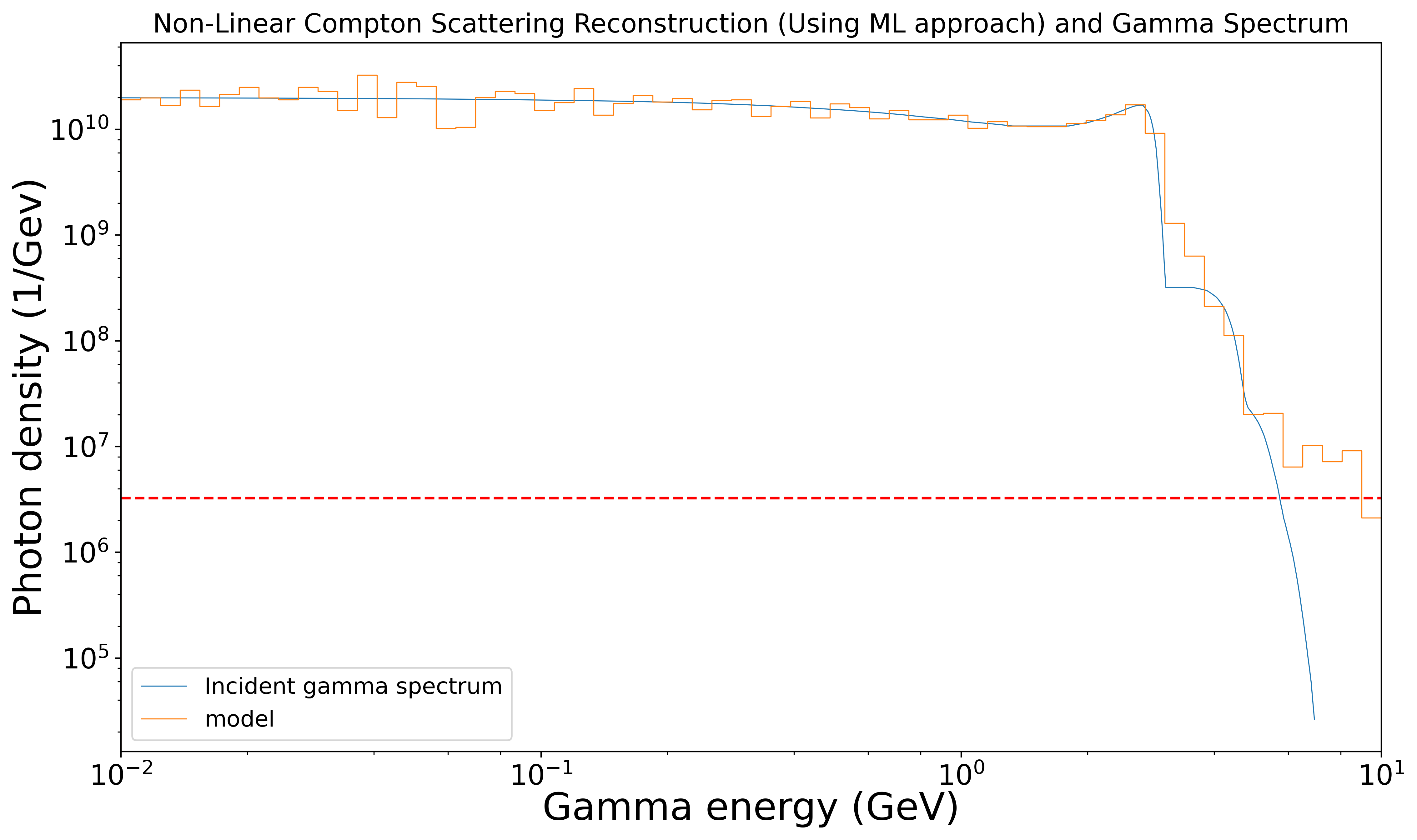}} \\
  \subfloat[Quantum electrodynamics case]{\includegraphics[width=\columnwidth]{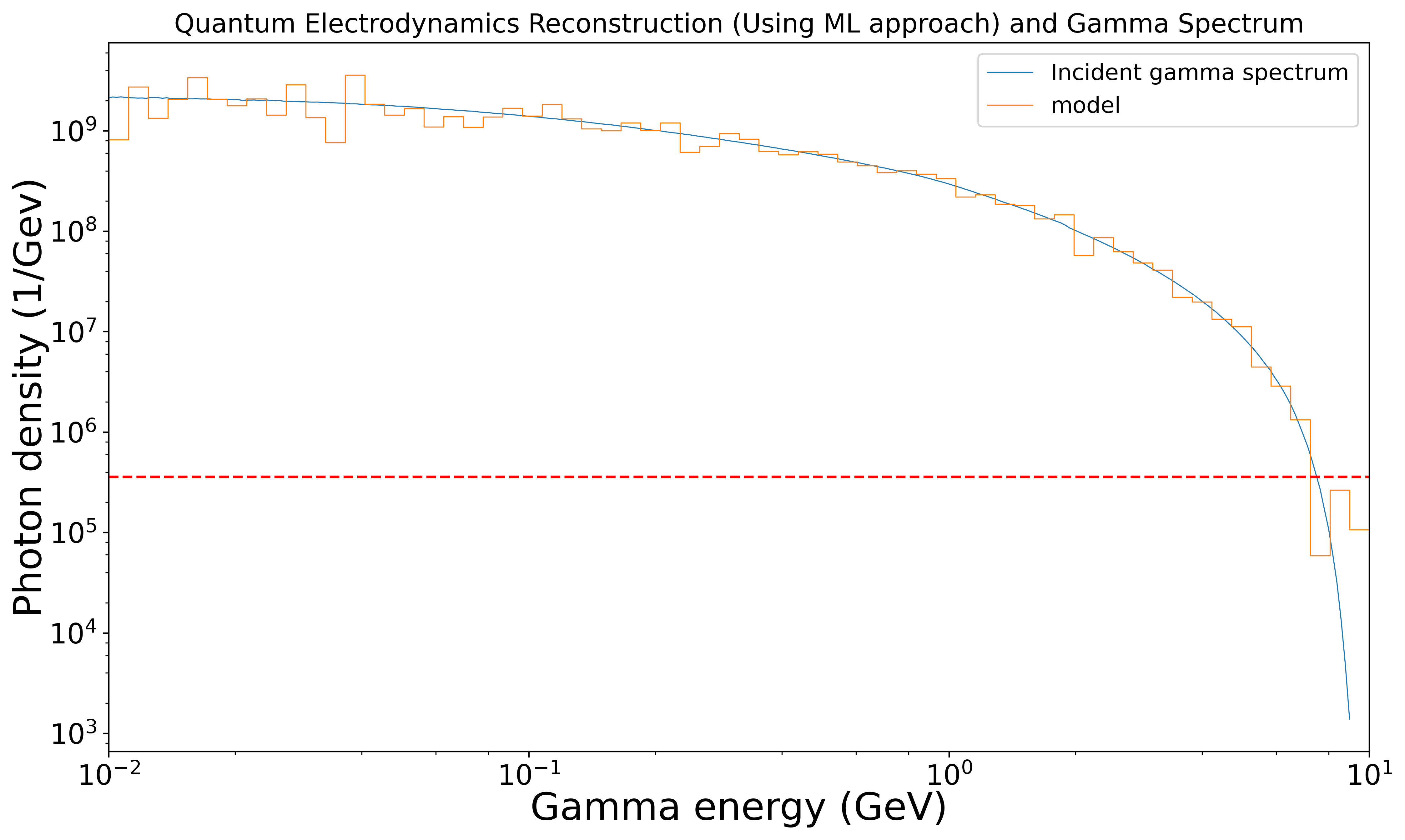}} \\
  \subfloat[Filamentation case]{\includegraphics[width=\columnwidth]{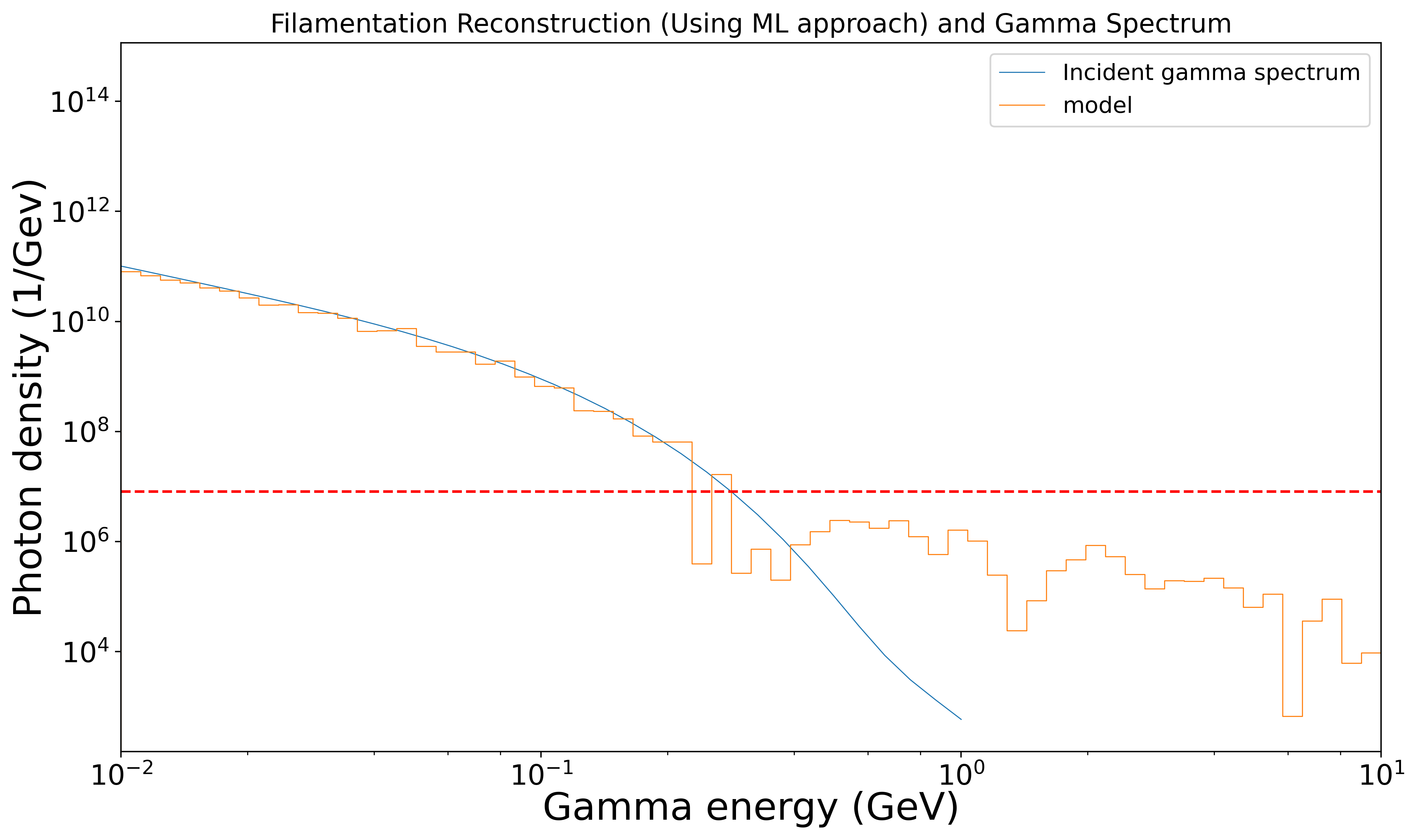}}
  \caption{Smooth reconstructed gamma distributions using machine learning.}
  \label{fig:3}
\end{figure}

An additional test on the model's performance was conducted to see its viability in smooth cases that more closely represented experimental results. Fig. \ref{fig:3} summarizes the model's performance. 

\begin{figure}[t]
  \centering
   {\includegraphics[width=0.47\textwidth]{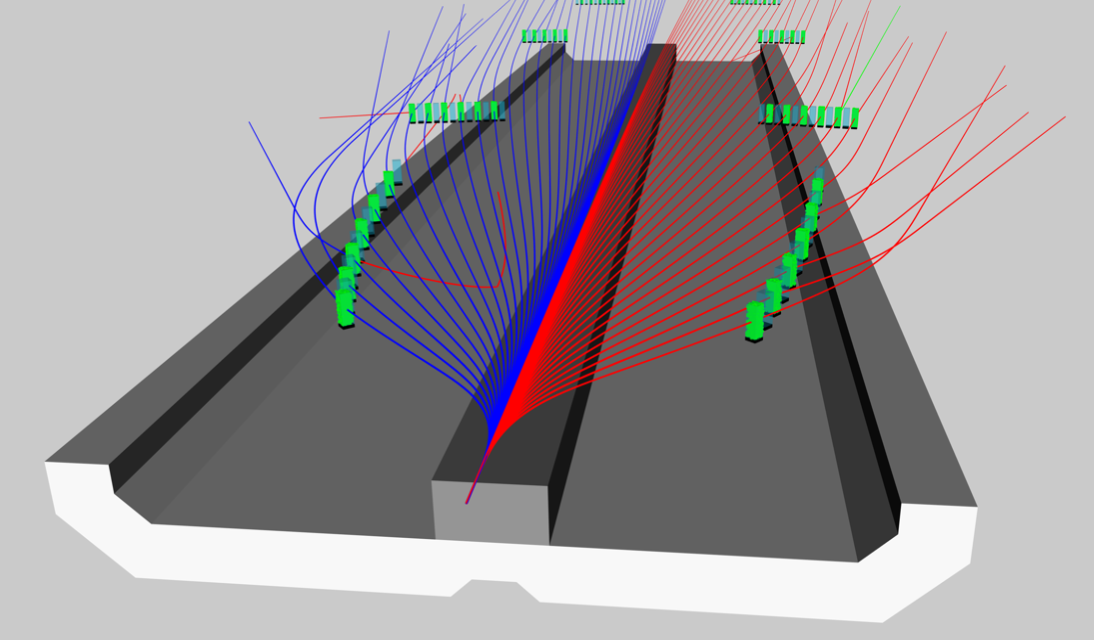}}
  \caption{Design overview. Not all features, particularly the SiPM PCBs and associated cabling, are shown for clarity. Cutaway view of design trajectories. Electrons are shown in red; positrons are shown in blue, and photons are shown in green. Energies span 10 MeV through 10 GeV. Note that the Cherenkov cells are oriented normally to incoming design trajectories.}
  \label{fig:pedro}
\end{figure}

In Fig. \ref{fig:3}, the model predicts the correct distribution pattern between bins 1 and 30 (approximately corresponding to a range of 10-250 MeV). In the higher energy bins, the model's predicted photons fall within the spectrometer's dynamic range in the nonlinear Compton scattering (Fig. \ref{fig:3}(a)) and quantum electrodynamics (Fig. \ref{fig:3}(b)) cases. In these two cases, the model follows the entire pattern of the reconstructed distribution, consistently staying within the same order of magnitude as the gamma distribution except at the highest energies, which are not of high physical interest.

In the filamentation case Fig. \ref{fig:3}(c), the model predicts the distribution to be higher in frequency than the actual distribution in the higher energy bins. This error occurs near the noise floor of the detector. The model is closely aligned with the gamma distribution in the 10-250 MeV range (bins 0-30).

\subsection{Maximum likelihood estimation and machine learning (hybrid) for PEDRO analysis }

As noted above, MLE is an algorithm that, given an initial guess for the solution of an equation, iteratively converges to the nearest solution that is the most probable. In the present context, this method converts the issue of calculating the original energy distribution based on the PEDRO output from an analytical problem to a statistical one to estimate a series of parameters \cite{MYUNG200390}. 

While the MLE algorithm does not require training like the ML model, it does require an initial guess that is sufficiently close to the proper solution to Eq. \ref{eq:pedro}. Given the enormous potential for variety in \textit{y}-vectors and incoming energy distributions, it is likely that a random guess will not reliably converge to the desired result. However, given the utility of machine learning, it is possible to provide a customized guess for every \textit{y}-vector: a guess provided by the  ML model. This combination serves as a hybrid approach to recovering the original energy distribution instead of pure ML (or QR decomposition, as will be seen in the next section). 

Using the ML model's guess, which may be close to the true distribution, the MLE algorithm may also reliably converge to the true distribution, as the issue of incorrectly identified convergent points is suppressed. 

\begin{figure}[t]
  \centering
   \subfloat[Mono-energetic case]{\includegraphics[width=\columnwidth]{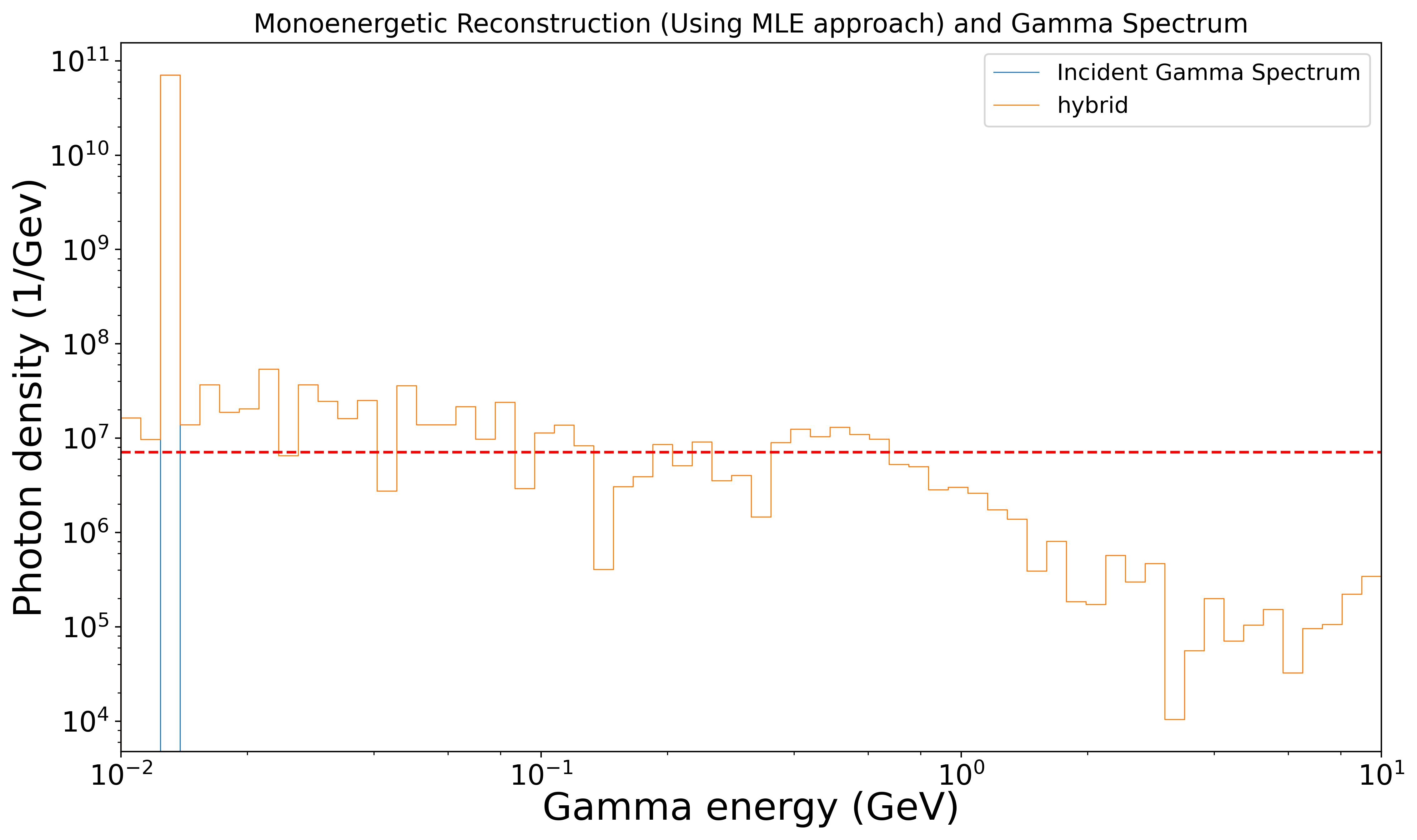}}\\
    \subfloat[Arbitrary Discrete case]{\includegraphics[width=\columnwidth]{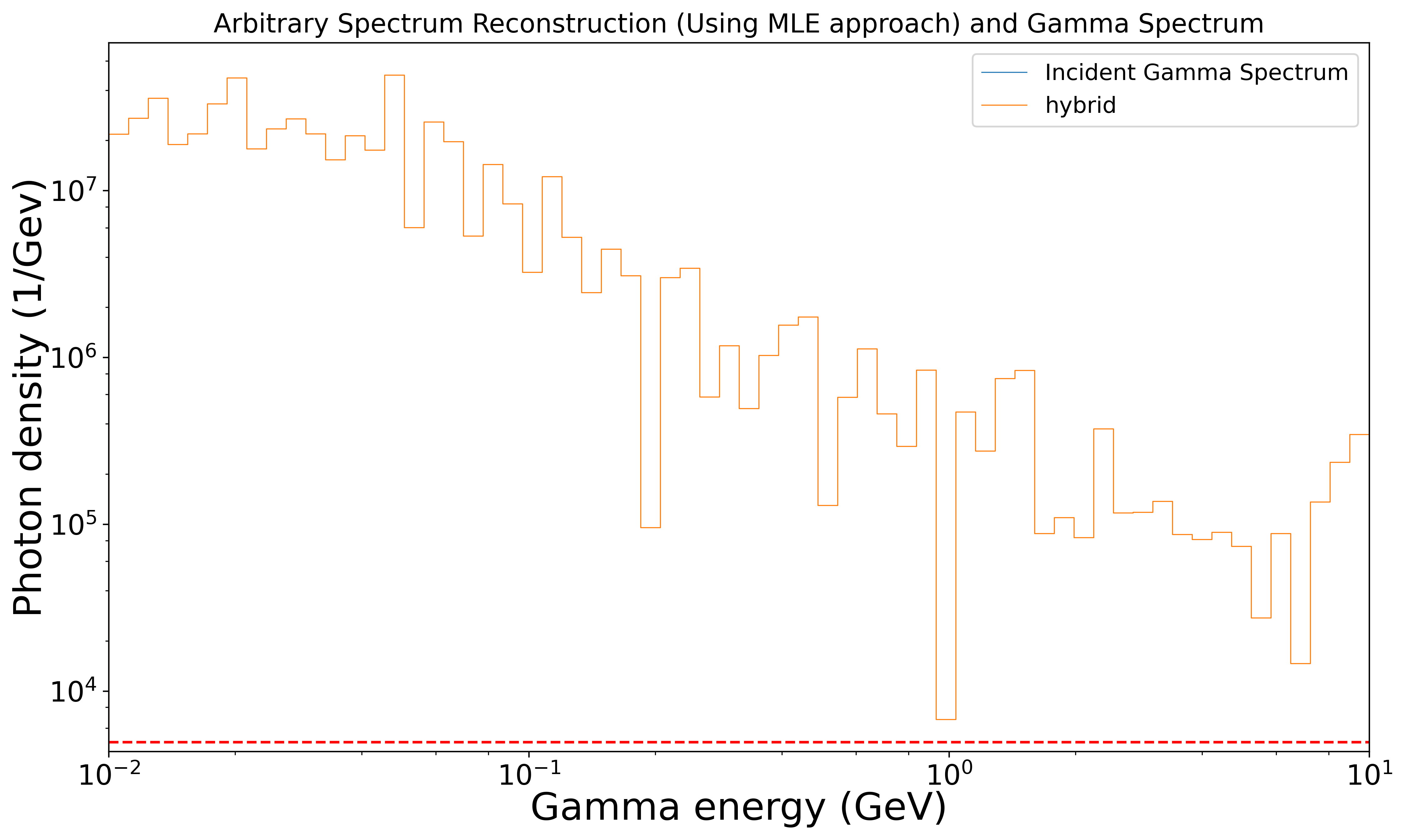}}
  \caption{Discrete reconstructed gamma energy distributions using the hybrid approach}
  \label{fig:4}
\end{figure}

In the discrete spectral cases Fig. \ref{fig:4}, the hybrid approach correctly predicts the location and heights of the peaks associated with the energy ranges of the true photon spectra. According to the MLE prediction, all other energy bins contain anywhere ranging from 1 to $5 \times 10^{5}$ photons. The predicted frequency of photons vanishes above the 10 GeV range after the rightmost peak in both cases, as must be the case on physical grounds. In the arbitrary discrete case, the hybrid approach accurately reconstructs the spectrum, with some slight deviations in the 250-670 MeV range (bins 30-40). This reconstruction remains within the same order of magnitude across all energy ranges, with quality consistent with expected experimental noise. 

\begin{figure}[!bht]
  \centering
    \subfloat[Non-Linear Compton scattering case]{\includegraphics[width=\columnwidth]{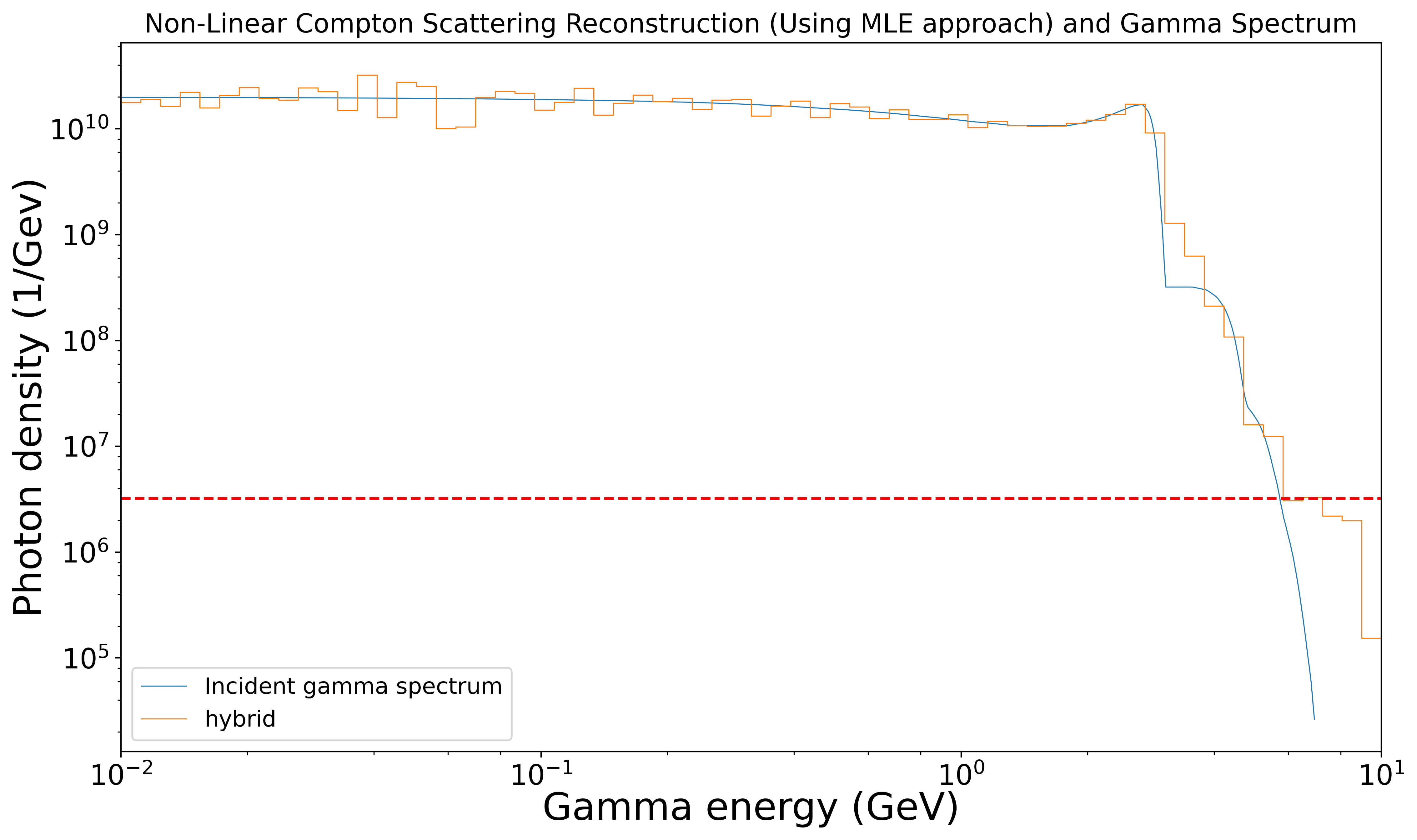}}\\
    \subfloat[Quantum electrodynamics case]{\includegraphics[width=\columnwidth]{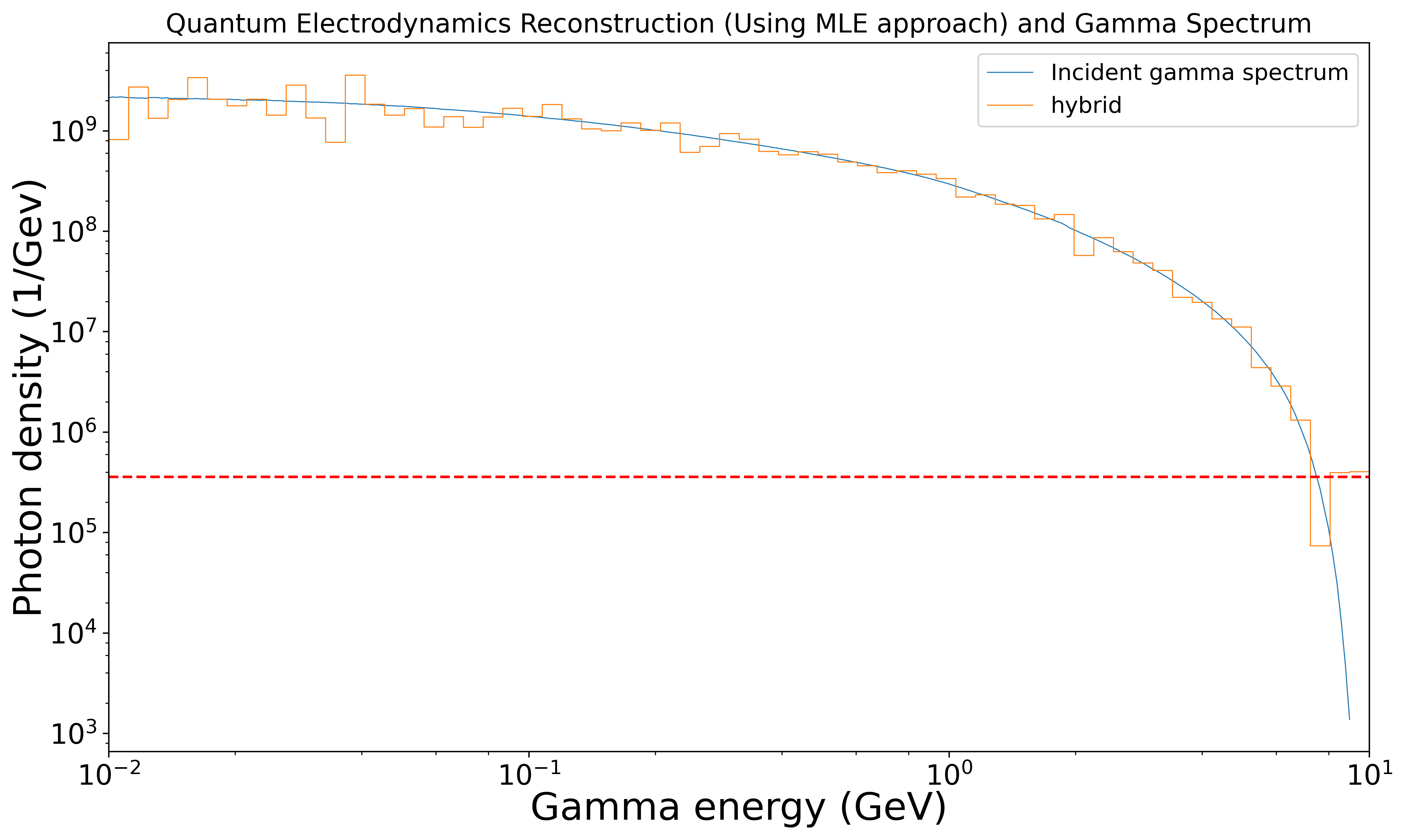}}\\
    \subfloat[Filamentation case]{\includegraphics[width=\columnwidth]{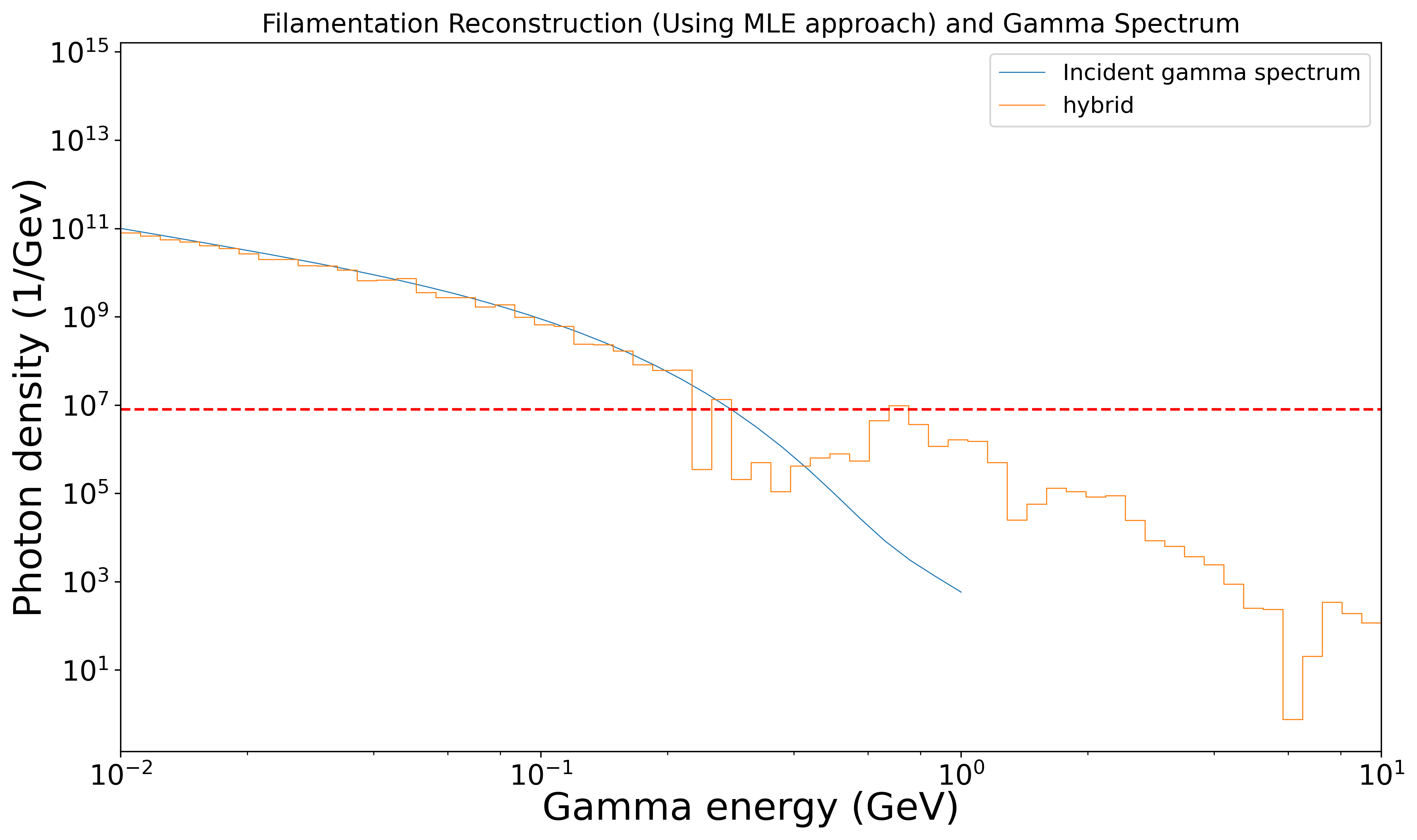}}
  \caption{Smooth reconstructed gamma distributions using the hybrid approach}
  \label{fig:5}
\end{figure}

In the smooth cases from Fig.~\ref{fig:5}, the hybrid approach closely follows the pattern of the gamma distribution across all of the energy ranges (with a deviation from the filamentation case above 250 MeV). In the nonlinear Compton scattering and quantum electrodynamics cases, the reconstructions consistently fall within the same order of magnitude as the gamma distribution and follow a similar pattern. 

In the filamentation case, the hybrid approach slightly underestimates the photon frequencies between 0-20 MeV, where the detector is not physically optimized but closely follows the distribution. Compared to the ML model's performance, the hybrid approach regularly fixes some deviations and smoothens the reconstruction, improving the accuracy. This also applies to the other two cases. 

\subsection{QR decomposition for PEDRO analysis}

The computational analysis goal is to invert the response matrix \textit{R} effectively so, given a PEDRO spectrum, the originating \textit{x}-vector may be recovered. This recovery method required no additional information beyond the matrix to be decomposed (in this case, \textit{R}). 

\begin{equation}
\label{eq:QReqn}
    \hat{R} = \hat{Q} * \hat{S}
\end{equation}

In this decomposition, $\hat{Q}$ is a square orthonormal matrix, meaning each of the column vectors $Q_i$ are orthogonal to each other and have a generalized unity length. This yields a convenient property that the transpose of $\hat{Q}$, $\hat{Q}^T$, is its inverse matrix, as shown below:

\begin{equation}
    \hat{Q} * \hat{Q}^T 
    =
   \begin{bmatrix} 
        \vdots & \vdots & \cdots & \vdots \\
         Q_1 & Q_2 & \cdots & Q_{64} \\
        \vdots & \vdots & \cdots & \vdots \\
    \end{bmatrix}
    \times
    \begin{bmatrix} 
        \cdots & Q_1 & \cdots \\
        \cdots & Q_2 & \cdots \\
        \vdots & \vdots & \vdots \\
        \cdots & Q_{64} & \cdots \\
    \end{bmatrix}
\end{equation}

\begin{equation}
     =
    \begin{bmatrix} 
        <Q_1,Q_1> & \cdots & <Q_1,Q_{64}> \\
        <Q_2,Q_1> &  \cdots & \vdots \\
        \vdots & \ddots & \vdots \\
        <Q_{64},Q_1> & \cdots & <Q_{64},Q{64}> \\ 
    \end{bmatrix}
\end{equation}

\begin{equation}
    =
    \begin{bmatrix} 
        1 & 0 & \cdots & 0 \\
        \vdots & \vdots & \ddots & \vdots \\
        0 & 0 & \cdots & 1 \\ 
    \end{bmatrix}
\end{equation}

As such, using this property and substituting the decomposition into the matrix equation that models the spectrometer's response, the equation transforms into:

\begin{equation}
    y = \hat{R}x = (\hat{Q}\hat{S})*x
    => \hat{Q}^Ty = \hat{Q}^T\hat{Q} * \hat{S}x
\end{equation}

\begin{equation} \label{eq:optimizer}
    => \hat{S}x = \hat{Q}^Ty = b
\end{equation}

Since, for a given response or \textit{y}-vector, the term $\hat{Q}^Ty$ will be constant, it will be referred to as the \textit{b}-vector throughout the remainder of this discussion. Under these assumptions, the method for finding the \textit{x}-vector is reduced to determining which solution will minimize the following function:

\begin{equation} 
    f(x) = ||\hat{S}x - b||^2
\end{equation}

Minimizing $f(x)$ will result in a solution for Eq.~\ref{eq:optimizer}, meaning using the least-squares optimization algorithm will provide the most likely energy distribution. Before implementing this algorithm, the matrix $\hat{S}$ values must be accounted for to determine whether the function is strongly convex. Strong convexity means the function is guaranteed to have a unique global minimum, which ensures the least-squares algorithm will converge to only one possible answer for a given \textit{b}-vector. 

To confirm strong convexity, the second derivative of $f(x)$ must generate a positive definite matrix, which is equivalent to stating the representation must be a symmetric matrix with strictly positive eigenvalues (all of the eigenvalues $\lambda_i > \;$0). 

\begin{equation}
    \frac{df(x)}{dx} = 2\hat{S}^T(\hat{S}x - b) = 2\hat{S}^T\hat{S}x - 2\hat{S}^Tb
\end{equation}

\begin{equation}
    => \frac{d^2f(x)}{dx^2} = 2\hat{S}^T\hat{S}
\end{equation}

Since $\frac{d^2f(x)}{dx^2}$ is equal to the product of a matrix with its transpose, it follows that:

\begin{equation}
\label{eq:Symeqn}
    (2\hat{S}^T\hat{S})^T = 2(\hat{S}^T)(\hat{S}^T)^T = 2\hat{S}^T\hat{S}
\end{equation}

\begin{figure}[t]
   \centering
     \subfloat[Mono-energetic case]{\includegraphics[width=\columnwidth]{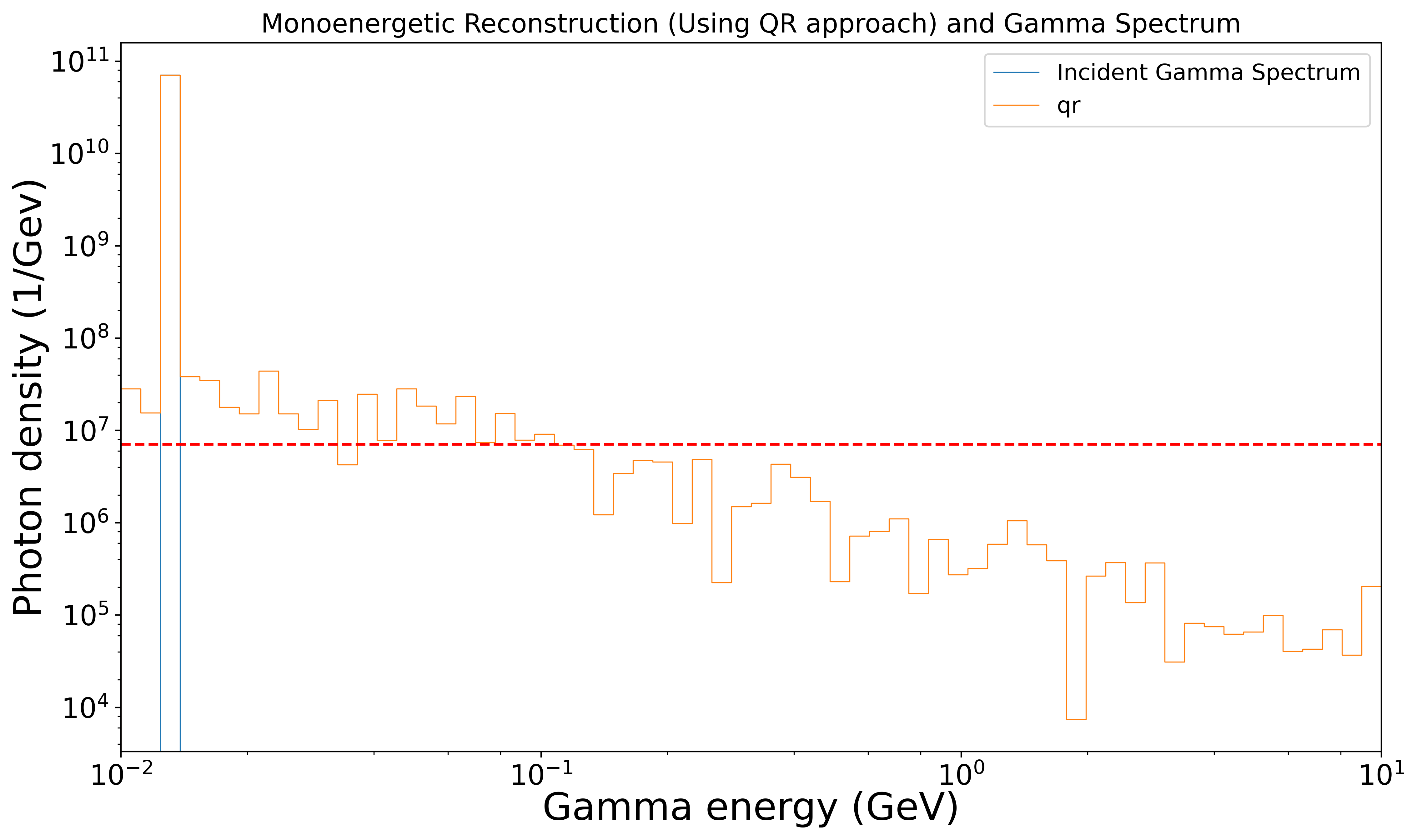}}\\
      \subfloat[Arbitrary discrete case]{\includegraphics[width=\columnwidth]{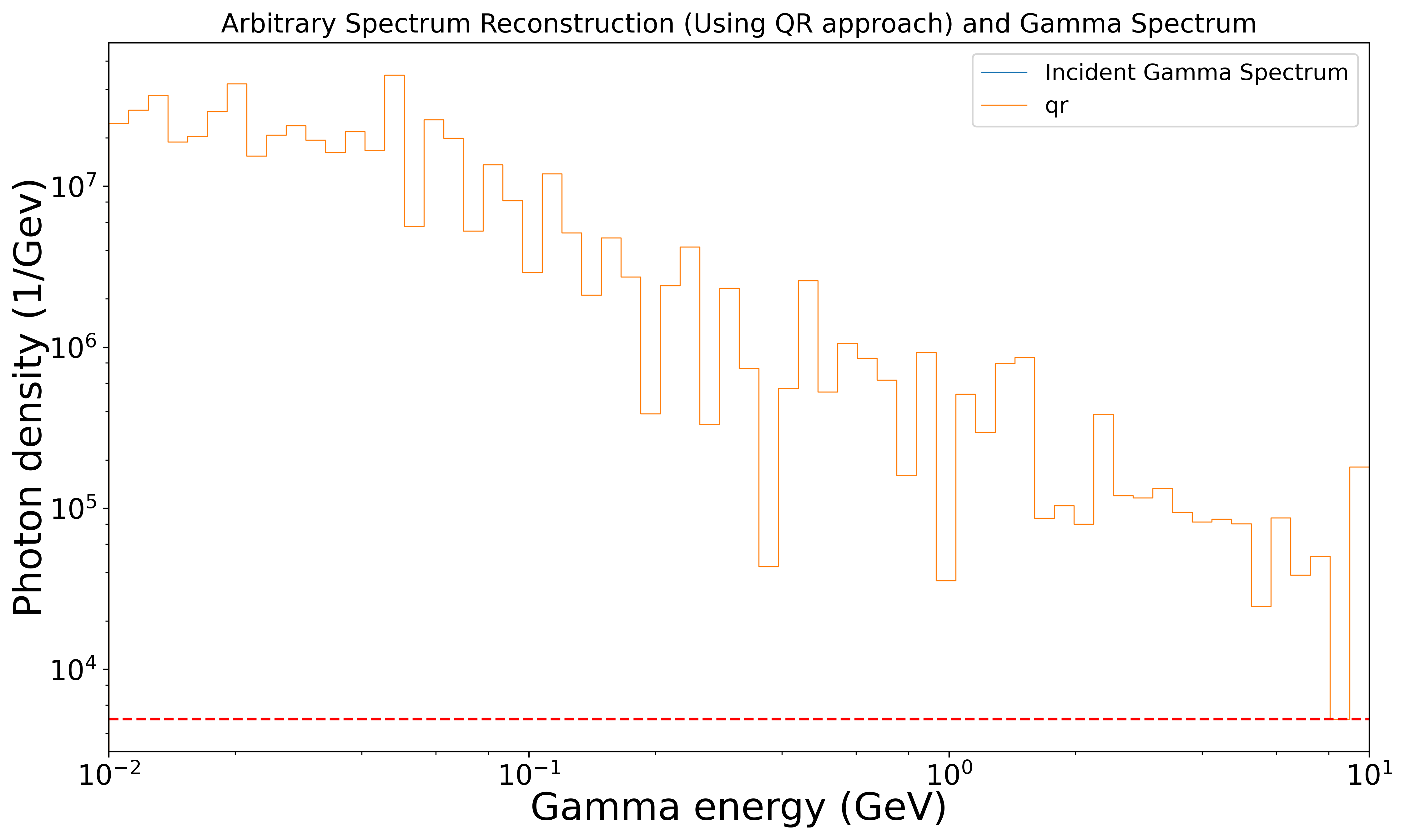}}
   \caption{Discrete reconstructed gamma energy distributions using QR decomposition}
  \label{fig:6}
\end{figure}

Using Eq.~\ref{eq:Symeqn}, $\frac{d^2f(x)}{dx^2}$ forms a symmetric matrix, meaning the only test that remains to determine the definite positive nature of $\frac{d^2f(x)}{dx^2}$ is to determine its eigenvalues. The Python NumPy library contains a function that takes a matrix as an argument and returns an array of its eigenvalues. Applying this function to $\frac{d^2f(x)}{dx^2}$ demonstrated that it indeed does have only strictly positive eigenvalues.

Thus, using the least squares optimization algorithm will avoid the issue of converging on potentially multiple solutions. After generating the QR decomposition, like the other approaches, this algorithm was tested against both discrete cases and more continuous cases.

Examining the predicted distributions for spectra generated from mono-energetic and arbitrary discrete distributions in Fig. \ref{fig:6}, the QR decomposition method predicts the peaks and their locations in photon frequency exceedingly well. In the arbitrary discrete case, the reconstruction shows no deviations from the gamma distribution. Despite being a purely mathematical reconstruction with no adjustments, QR decomposition performs similarly to the ML and hybrid methods in reconstructing the mono-energetic case. In addition, the noise floor in the reconstruction is extremely low, with frequency predictions below one per bin in the region away from the true distribution.

\begin{figure}[t]
  \centering
    \subfloat[Non-Linear Compton scattering case]{\includegraphics[width=\columnwidth]{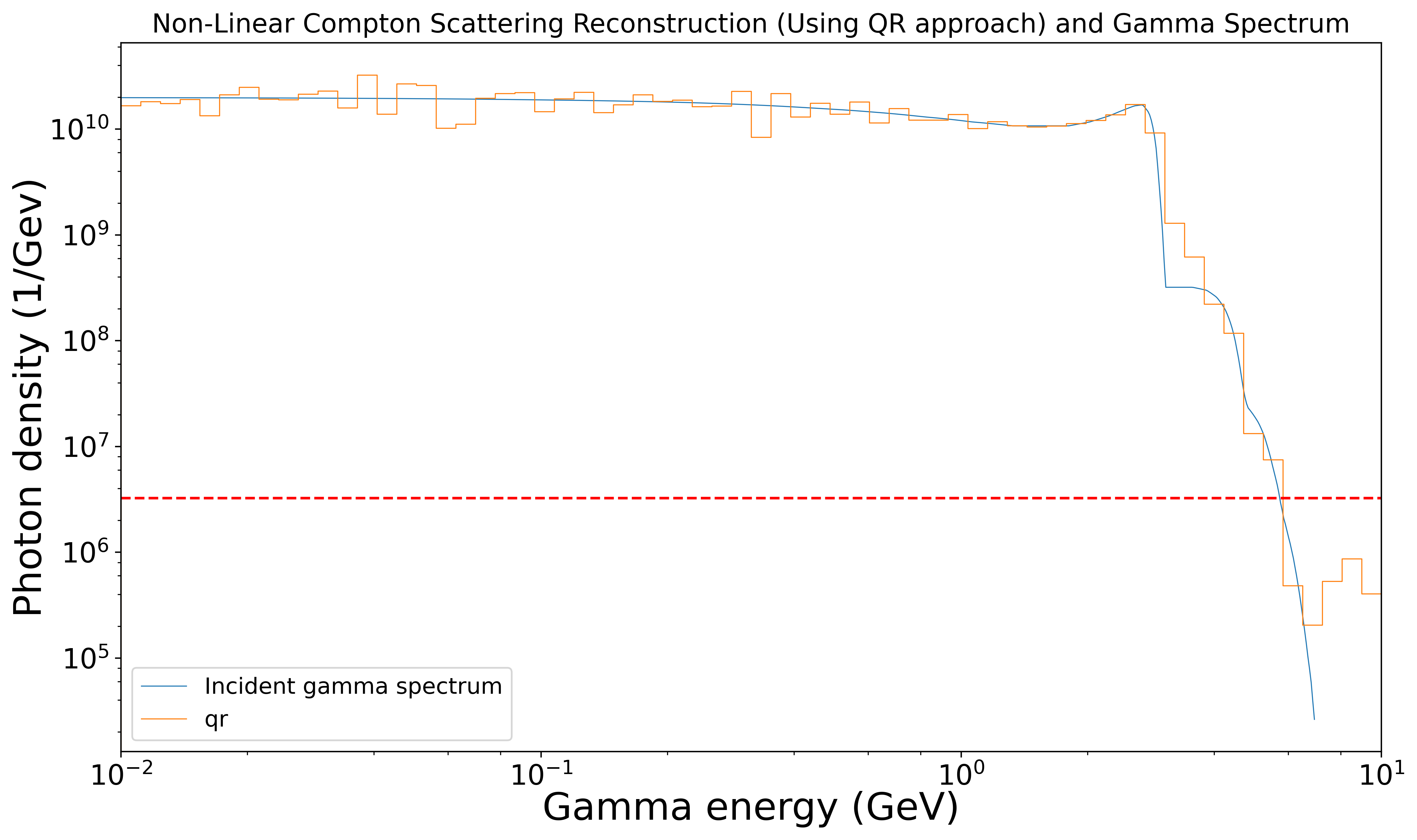}}\\
    \subfloat[Quantum electrodynamics case]{\includegraphics[width=\columnwidth]{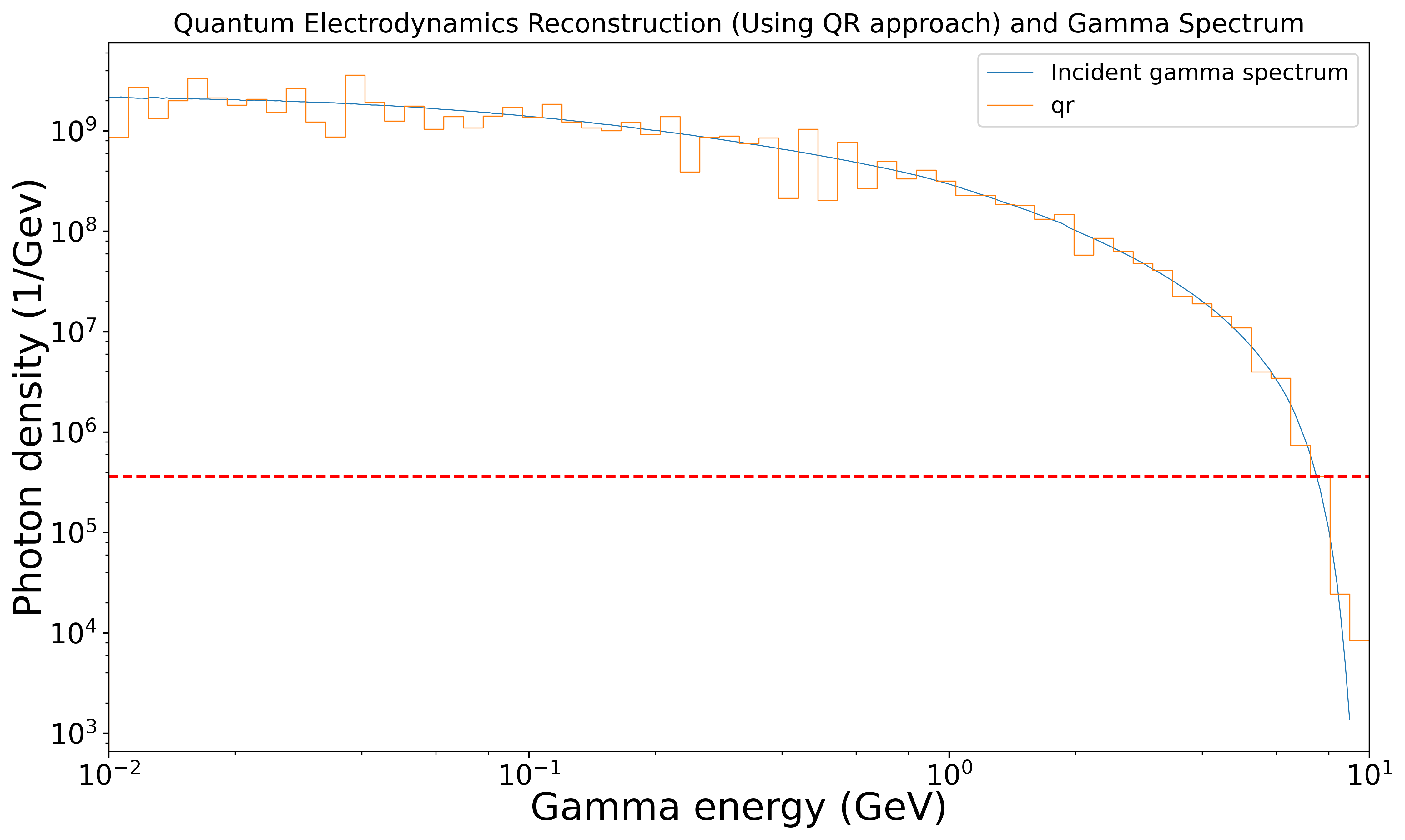}}\\
    \subfloat[Filamentation case]{\includegraphics[width=\columnwidth]{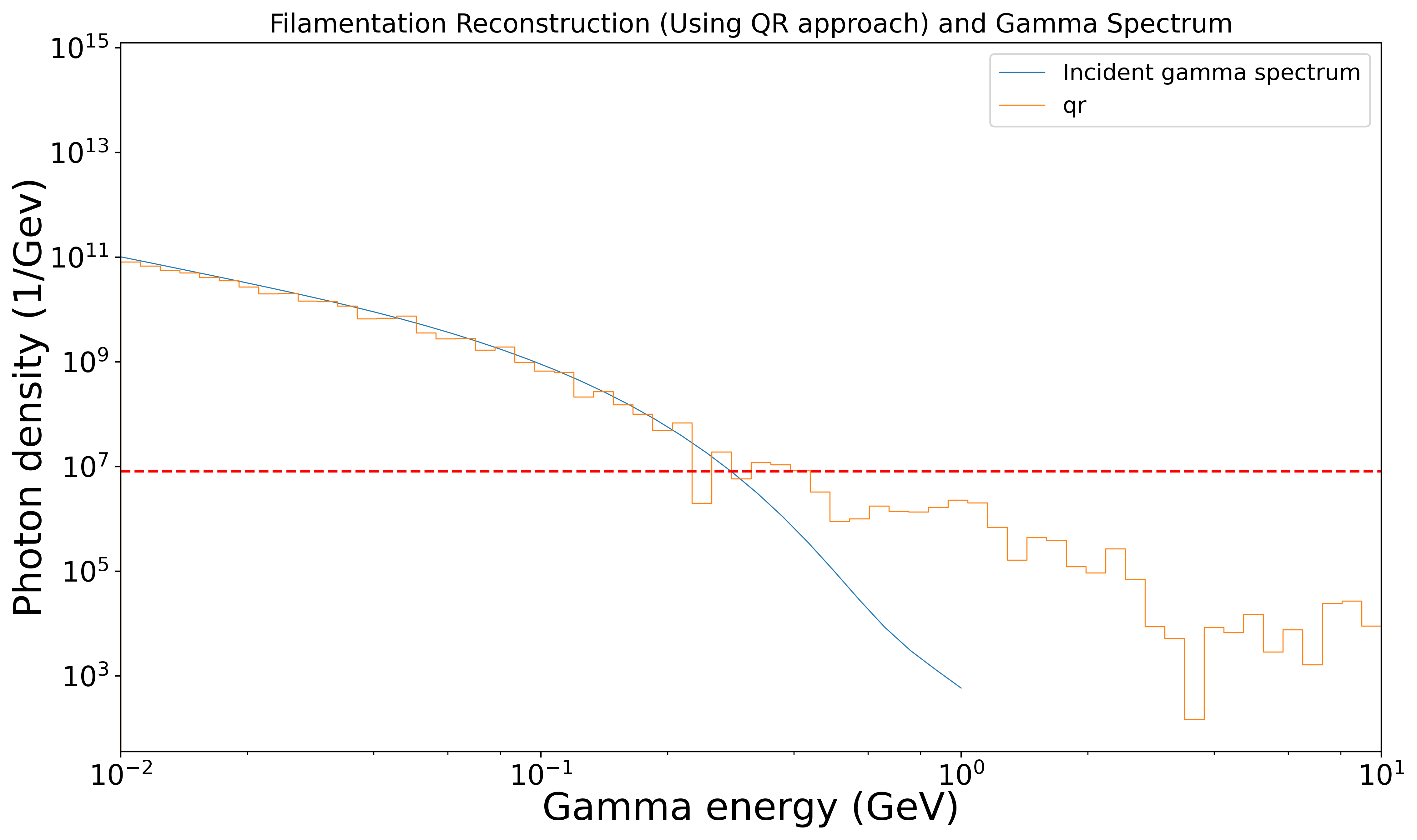}}
  \caption{Smooth reconstructed gamma distributions using QR decomposition}
  \label{fig:7}
\end{figure}

In all three of the smooth distributions (see Fig. \ref{fig:7}), the simulated and all reconstructed spectra follow extremely close to each other. Within a single iteration, QR decomposition generates distributions with an accuracy similar to that of the other two approaches. Across all energy bins, the QR decomposition method predicts the correct pattern and frequencies of photons, with the standard exception of the range above roughly 250 MeV in the filamentation case. 

\subsection{Discussion of PEDRO analysis}
\label{sec:discussion}

\begin{figure}[t]
    \centering
    \includegraphics[width=\columnwidth]{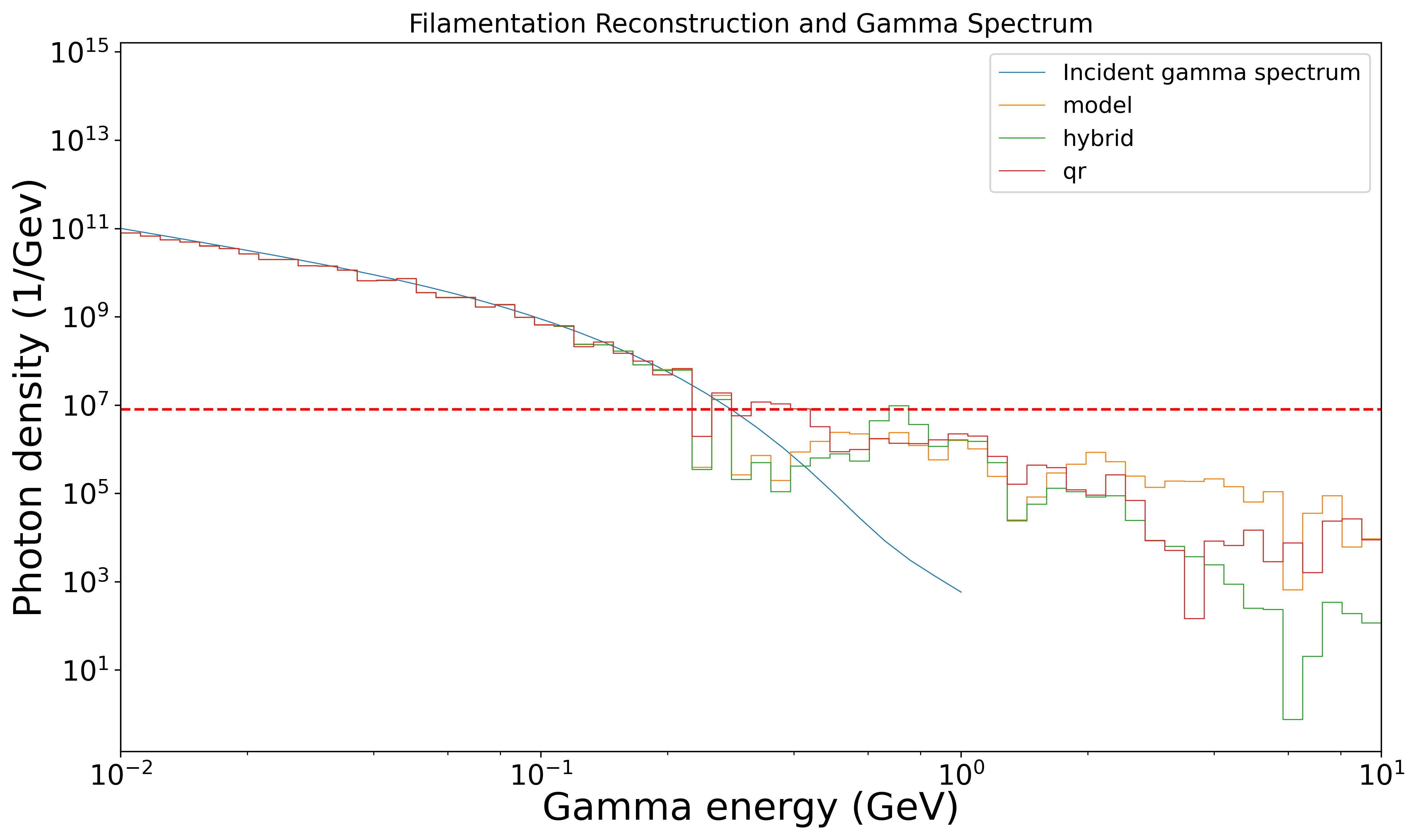}   \caption{Reconstruction of filamentation case using three different methods.}
    \label{fig:All}
\end{figure}

Examining the results above, all of these approaches provide similar performance levels, with the QR decomposition method requiring the least amount of calculation time to implement, given its non-iterative algorithm in conjunction with least squares optimization. In addition, the ML and ML-MLE hybrid approaches present highly similar performance levels, with the hybrid approach providing somewhat smoother solutions. All three methods are presented in the quantum electrodynamics test case together in Fig. \ref{fig:All}.

The mathematical justification of implementing QR decomposition also helped improve the legitimacy of its solution since the solution was guaranteed to be unique. However, while the QR decomposition worked well in this scenario, it is not guaranteed to generalize to other reconstruction analyses with a linear representation. With this caveat in mind, the computational approach for gamma energy spectra in PEDRO is in practice best solved using QR decomposition.

Another point to consider is the presence of noise in actual physical experiments. While steps are continually being taken on the instrument side to reduce noise issues, any reconstruction method should remain robust in the face of noise. Given this requirement, and the similar performance of all three methods in the experimental cases, the hybrid approach works best in reconstructing the gamma spectra from PEDRO spectrometer readings. 
\cite{Maanas,yadav2021_AAC, 2021yadav,zhuang1,Naranjo_1,naranjocompton,yadav_2022}

Similar reconstruction work has been conducted in the past, as noted in the case of the Compton Spectrometer, which dealt with a nonlinear system of reactions to incoming particles. Given that PEDRO's responses were entirely linear (not with standing the standard experimental noise), the selected methods of analysis utilized different assumptions than the CPT software, thus resulting in different neural network architecture and conclusions. 

The work done with PEDRO assumed the data to be analyzed derived from the single-shot case, where only one measurement would be taken for a given electron beam-laser interaction. However, in the multi-shot case, the same interaction can be repeatedly measured with different horizontal positions of the converter target wired to obtain double-differential spectra, yielding more information about these high-energy interactions. Further, we may look at sparsely populated regions of the spectrum, thus extending the dynamic range obtained. As such, a future direction for the work in the paper is to generalize its reconstruction methods to the more complicated but rewarding multi-shot case. 

\section{Outlook and future work}
\label{sec:conclusion}

This work demonstrates that both MLE and ML can effectively use betatron radiation data in implementation as a tool for extracting beam diagnostic information. Specifically, it may be employed to identify beam spot sizes, emittances, and energies. The ability to simultaneously predict multiple parameters with a single model is an attractive goal for future work. The discussed methods have possible applications for various PWFA experiments at FACET-II or other electron beam facilities where betatron radiation is the most robust and accessible experimental measurement capable of revealing fundamental aspects of the beam-plasma interaction.

To give a compelling example, the methods discussed here may also be used to explore the scenarios of the plasma-based terawatt attosecond (PAX) project, also planned at FACET-II. PAX focuses on that electron beams from PWFAs can generate few-cycle coherent tunable soft X-ray pulses with terawatt peak power and duration of tens of attoseconds. In addition, these beams will be shorter with better stability than state-of-the-art x-ray free electron lasers (XFELs) \cite{PAXclaudio}. The PAX scenario can be experimentally realized by employing two PWFA stages with different plasma densities. The first stage will create a beam with a linear chirp that can be removed using a chicane transformation, yielding a very high current, ultra-short (10's of a nanometer in length). This beam can be used to drive plasma wakefields in a very dense plasma \cite{pratikpax2022}, after focusing using a permanent magnitude quadrupole (PMQ) triplet to reduce its transverse size further. This beam is capable of driving TV/m-class wakefields in a very high-density plasma. 

The betatron radiation spectra emitted by the electron beam for the parameters given, particularly for densities over $10^{20}$ cm$^{-3}$ show significant new signatures. When PAX parameters at these densities are used in simulation, photons are produced well into the MeV range. The spectrum is broadened by the particular nonlinear focusing of the beam fields. Liénard–Wiechert field-based simulations are now underway at UCLA \cite{monikaprab, SakaiICS2017} to provide detailed theoretical predictions of the spectral shape. Measurement of the many MeV-scale double-differential spectra can be accomplished by use of the UCLA-developed spectrometers \cite{naranjocompton, Naranjo_1}.

In summary, the investigatino of the physics of critically important phenomena in the PWFA has been explored through sophisticated gamma-ray spectrometer output analyzed with advanced ML and MLE techniques. This has required a theoretical and computational approach undertaken in the particular experimental context of the FACET-II research program. The issues examined here provide for a rich physics study of beam parameters and spectrum reconstruction using the methods discussed in this paper. Successful exploration of the physical effects involved should give confidence in the physical understanding of PWFA and its experimental control. This work leads towards significant advancements in beam diagnostic techniques, which are of necessity very different in a PWFA-based future linear collider than in the current state-art-of-the-art accelerators.

The spectral and angular photon spectral yield can then be combined to provide an indirect measurement of the beam's phase space distribution while inside the plasma, which is critical to fully understanding—and ultimately optimizing—the beam dynamics inside a plasma wakefield accelerator, in order to understand high-brightness beam behavior for high energy physics applications. It may also be used to detect deviations from ideal focusing conditions in the plasma, such as an unmatched beam or one with a significant off-axis tail. Further, one should contemplate effect that might arise from ion collapse\cite{ioncollapse}, a potentially severe problem for linear collider beams in a PWFA. Ion collapse has been recently understood to be the collisionless relaxation of the beam-ion system towards a quasi-equilibrium of a three-species (ions, plasma, and beam electrons) system. The non-linearity of the fields during ion collapse causes filamentation of the beam's phase space, which may grow the emittance\cite{weiming}. Betatron radiation signals may also enable the detection of beam breakup and chromatically driven emittance growth.

This type of measurement, in which the double-differential radiation spectrum (DDS) is obtained on a single-shot basis, has already been undertaken by the UCLA team at the BNL ATF \cite{SakaiICS2017}. This photon spectral measurement concentrated on keV-class X-rays in the context of the nonlinear Compton scattering (ICS) experiment. The use of the DDS tool revealed significant physics detail concerning the electrodynamics of this system. 

The methods discussed above can transform these measurements into tools for beam diagnostics for a host of betatron radiation-producing experiments and for Compton systems such as the nonlinear strong-field QED experimental. Indeed, we have seen that the reconstruction of the primary gamma spectrum from pair data is robust for various methods employed.  

These physics goals discussed above are all found in the wide-ranging FACET-II program and beyond to other present and future laboratory facilities. As radiation-based diagnostics are key, robust signatures of the physics involved in these experiments, the introduction of modern methods for analyzing such data is an essential component of the FACET-II effort. This paper has provided a data analysis study for the advanced gamma spectroscopy instrumentation at FACET-II. It shows that these methods, in tandem with advanced spectrometers, are promising for extracting the needed physics insights from experiments ranging from PWFA to strong field QED. 

\section{Acknowledgements}

This work was performed with the support of the US Department of Energy, Division of High Energy Physics, under Contract No. DE-SC0009914, NSF PHY-1549132 Center from Bright Beams, DARPA under Contract N.HR001120C007 and the STFC Liverpool Centre for Doctoral Training on Data Intensive Science (LIV.DAT) under grant agreement ST/P006752/1.

\bibliography{main.bib}

\end{document}